\documentclass[conference]{IEEEtran}
\usepackage{graphicx}
\usepackage{booktabs} 
\usepackage{etoolbox}
\usepackage{amsmath}
\usepackage{amsthm}
\usepackage[noend]{algcompatible}
\usepackage{algorithm}
\usepackage{epstopdf}
\usepackage{booktabs}
\usepackage{multirow}
\usepackage{hyperref}
\usepackage{array}
\newcolumntype{P}[1]{>{\centering\arraybackslash}p{#1}}
\usepackage{cite}
\usepackage{amsmath,amssymb,amsfonts}
\usepackage{textcomp}
\usepackage{xcolor}
\def\BibTeX{{\rm B\kern-.05em{\sc i\kern-.025em b}\kern-.08em
    T\kern-.1667em\lower.7ex\hbox{E}\kern-.125emX}}

\newtheorem{problem}{Problem}

\newtheorem{definition}{Definition}

\newtheorem{theorem}{Theorem}
\newtheorem{lemma}{Lemma}

\def\techReport{}
\begin{document}
\title{Finding Average Regret Ratio Minimizing Set in Database}
\author{\IEEEauthorblockN{Sepanta Zeighami}
\IEEEauthorblockA{\textit{Hong Kong University of Science and Technology} \\
Kowloon, Hong Kong \\
szeighami@cse.ust.hk}
\and
\IEEEauthorblockN{Raymond Chi-Wing Wong}
\IEEEauthorblockA{\textit{Hong Kong University of Science and Technology} \\
Kowloon, Hong Kong \\
szeighami@cse.ust.hk}
}

\maketitle

\begin{sloppy}
\begin{abstract}
Selecting a certain number of data points (or records) from a database which ``best'' satisfy users' expectations is a very
prevalent problem with many applications. One application is a hotel
booking website showing a certain number of hotels on a single page.
However, this problem is very challenging since the selected points should ``collectively'' satisfy the expectation
of all users. Showing a certain number of data points to a single user
could decrease the satisfaction of a user because the user may not be able to
see his/her favorite point which could be found
in the original database. In this paper, we would like to
find a set of $k$ points such that on average, the satisfaction (ratio) of
a user is maximized. This problem takes into account the probability distribution of the users and considers the satisfaction (ratio) of
\emph{all} users, which is more reasonable in practice, compared with the existing studies
that only consider the \emph{worst-case} satisfaction (ratio)
of the users, which may not reflect the whole population and is not useful
in some applications. Motivated by this, in this paper, we 
propose
algorithms for this problem. Finally, we conducted experiments
to show the effectiveness and the efficiency of the algorithms.
\end{abstract}

\section{Introduction}
\label{sec:intro}

Selecting a certain number of data points (or records) from a database in order to ``best'' satisfy users' expectations  is a prevalent problem with many applications, from recommender systems to search engines. In many situations, the users of such an application is \textit{anonymous}, that is, s/he not registered on the website or has not logged into his/her account. No personal information is available regarding the specific preferences of an anonymous user. Thus, only general information about the users (possibly refined by user's location) can be utilized to select the data points for the anonymous user. To see the problem more clearly, consider the following example. 

Consider using a website such as booking.com to book a hotel room in London. The website offers more than 6,000 properties in London. The users can perform queries based on the location, star rating and etc. to reduce the number of properties to an order of 100 properties that match their criteria. The users need the website to show them a smaller number of hotels for selection, as it is infeasible for them to go through all the matching hotels. Users can perform bookings without having an account and thus no personal information is available about the user's preferences when the items are shown to such users.

In addition to the case that the preferences of the users looking for a data point is not available, the users themselves may not be certain of their preferences. For instance, when looking for a restaurant on Yelp, a user might not know what type of food s/he might want to eat. Yet, there may be many restaurants in the location and the price range specified by the user. It would help the users' decision-making if only a small number of selected restaurants were shown to the user. Similar scenarios hold for apartment or car rental websites and other online shopping platforms, or when searching for a video without signing into a website such as youtube.com. 

More specifically, when a number of points are selected for a user, no specific information may be known about the user. The main challenges in selecting a number of data points for such a user are that, firstly, very little is known about the user's preferences: the websites usually do not have an interface to ask the users to input their preferences (beyond merely refining a query) and even if they did, the users might not be willing or capable to provide their exact preferences. Secondly, the user will have a low \emph{satisfaction} level if s/he cannot see his/her favorite data point among the points shown to the user. Thus, an accurate formulation of the level of satisfaction of the user is required for us to be able to select data points that better satisfy the user, without knowing the user's preferences.

In the literature~\cite{topk:Ilyas, topk:Papadopoulos} of top-$k$ query processing, the concept of utility functions, which are real-valued functions defined over the set of points in the dataset, are used to quantify users preferences and how much a user ``likes'' a \textit{single} data point. In our setting, we consider the case that when selecting a set of points for a user, the user's utility function is unknown. In the absence of any knowledge about the user's utility function, we cannot select points that maximize the user's utility function. Instead, we consider the problem of selecting a set of items that are \textit{expected} to \textit{satisfy} the user the most.

To tackle this problem, first we need a formulation for the satisfaction of a user when a \textit{set of points} is shown to a user (as opposed to the user's satisfaction from a \textit{single point}, which is measured by the user's utility function). Recently, The concept of \emph{regret ratio} and the $k$-regret problem  \cite{ANZ+17,AKS+17,CLW+17,KBL15,NLS+12,chester:regret,regret:Peng,Nanongkai:regret} has been proposed to measure how well a \textit{set of points} satisfies a user compared with when the user has seen the entire database.  Specifically, when a user is presented with a \textit{set of points}, the \emph{satisfaction} of the user from the set of points is defined as the maximum value of the utility function of the user among all the points in the set. The \emph{regret} of a user is the difference between the satisfaction of the user when s/he is shown a set of points $S$ compared with when s/he is shown his/her favourite point in the database (that is compared with when his/her satisfaction is at maximum possible). The \emph{regret ratio} of a user is his/her regret normalized by his/her satisfaction from his/her favourite point in the database, which allows us to compare the regret ratio of different users. 

Regret ratio of a user can be considered an accurate measure to quantify how well a set of points are selected for the user. However, in our problem setting, because we do not know the user's utility function, we cannot select a set of points for the user in order to minimize the user's regret ratio. Instead, we can focus on selecting a set of point that minimize the user's regret ratio on expectation (i.e., on average), based on the probability distribution of the utility functions. 

Note that the probability distribution of the utility functions can be modeled using the information available about other users and their preferences based on their history and their feedback provided in online ratings and reviews, which is extensively studied in data mining and machine learning communities \cite{CC+15, SF+07, recommender1, LearningGaussianProcesses, LearningGaussianProcesses2}. Thus, we consider the scenario when the users' utility functions follow a given probability distribution.

In this paper, we study the following problem.
Given a database $D$ and a probability distribution $\Theta$ of utility functions, we want to find a set $S$ of $k$ points
in $D$ such that the expected (average) regret ratio of a user is the least.

All studies on $k$-regret queries so far have focused on minimizing the maximum regret ratio. Although minimizing maximum regret ratio provides a worst-case guarantee on the regret ratio of all the users, there are a number of issues it cannot address. Firstly, maximum regret ratio disregards the probability distribution of the utility functions and considers all the possible utility functions equally. However, it may be important to obtain a lower regret ratio for the utility functions that are more probable and occur more frequently (in the hotel booking example, it can be more important to have lower regret ratio for users who book hotels every month compared with the ones who book a hotel once a year). Maximum regret ratio cannot distinguish between users with different probabilities, and an improbable utility function might cause a set of points to have a high maximum regret ratio, while the set might be suitable for the frequent users.

Moreover, even when the utility functions are distributed uniformly, maximum regret ratio will still not be able to account for the distribution of the regret ratio among the users. That is, two sets can have the same maximum regret ratio, but the regret ratio of a large proportion of the users in one can be significantly smaller than the other. Yet, maximum regret ratio will not be able to differentiate between the two sets, while the average regret ratio of the two sets will be different. It is confirmed empirically by our experimental results on real datasets (see Section \ref{sec:exp}) that the vast majority of the users will have a lower regret ratio if we minimize average regret ratio instead of maximum regret ratio.

To be able to address these issues, we need to consider the distribution of the utility function as well as the regret ratio of all the users, which can be done by considering the \emph{average-case} scenario involving the expectation over all the users instead of merely optimizing the \emph{worst-case} scenario. 

In this paper, we formulated the problem called \emph{\underline{f}inding
the \underline{a}verage regret ratio \underline{m}inimizing set (FAM)}. Given a database $D$ and a probability distribution $\Theta$ of utility functions, we want to find a set $S$ of $k$ points in $D$ such that the expected \emph{regret ratio} of a user is the smallest.

Solving FAM is not trivial. We show that the problem is NP-hard, and solving it is computationally costly. Moreover, a straightforward implementation that enumerates all possible solutions is very inefficient and in our experimental results it takes more than 50 hours to select 5 points in a real dataset containing 100 points, which is not scalable.

The following shows our contributions.
Firstly, we are the first to study the FAM problem comprehensively.
We show that this problem is NP-hard.
The existing studies about $k$-regret queries cannot address the FAM problem and our experiments support that our result based on the average regret ratio
is better than the result generated by existing $k$-regret
queries.
Secondly, we observe two interesting properties
for FAM, namely the ``steepness'' and
the ``supermodularity''. Based on these properties,
we propose an approximate algorithm called {\scshape{Greedy-Shrink}} which can return the solution set efficiently with a theoretical error guarantee. As opposed to the existing $k$-regret query methods, {\scshape{Greedy-Shrink}} does not make any assumption on the form of the utility functions and does not depend on the dimensionality of the database. This allows our method to be efficiently applied under any scenario for the dataset and utility functions.

Thirdly, we provide an exact algorithm in the case of linear utility functions when the dimensionality of the database is equal to 2. We use a dynamic programming algorithm to solve the problem optimally.
Fourth, we conducted extensive experimental studies to show
that {\scshape{Greedy-Shrink}} has a good performance
in terms of the average regret ratio and the query time.

The rest of the paper is organized as follows.
Section~\ref{sec:def} formally defines our FAM problem.
Section~\ref{sec:algo} shows our proposed
algorithm.
Section~\ref{sec:twoDimAlgo} presents our dynamic programming algorithm in a 2-dimensional case.
Section~\ref{sec:exp} presents our experimental results.
Section~\ref{sec:rel} gives the related work.
Section~\ref{sec:con}
presents the conclusion and the future work.

\section{Problem Def. \& Property Def.}\label{sec:def}
\subsection{Problem Definition}
Given a database $D$ containing $n$ points, we want to select $k$ points that best satisfy users' expectations. We first need to define how users' feelings towards a point are captured. 

\begin{definition}[Utility function~\cite{Nanongkai:regret}]
A utility function $f$
is a mapping $f: D \rightarrow R_{\geq 0}$.
The \emph{utility} of
a point $p$ with respect to a user with utility function $f$ is denoted by $f(p)$. 
\end{definition}

\begin{table}
	\centering
	\begin{tabular}{|c|c|c|c|c|} \hline
		 &Holiday Inn &	Shangri la & Intercontinental & Hilton \\\hline
		Alex      & 0.9  & 0.7	     & 0.2	    & 0.4  \\\hline
		Jerry 	  & 0.6   &1 	 & 0.5	    & 0.2  \\\hline
		Tom   & 0.2 &0.6      & 0.3   	& 1	    \\\hline
		Sam    &   0.1       &0.2      & 1	    & 0.9  \\\hline
	\end{tabular}
	\vspace{0.1cm}
	\caption{Utility functions of the users for the hotels.}
	\label{tab:example:hotelUtil}
	\vspace{-0.6cm}
\end{table}

A utility function measures a user's satisfaction with a particular point. It can also be written as an $n$-dimensional vector where each attribute of the vector is the utility of the user from a point in the database. Using this representation, Table \ref{tab:example:hotelUtil} shows the utility functions of 4 different users regarding a dataset of 4 hotels (normalized by the largest utility value). Unless specified otherwise, we do not make any assumptions on the form of the utility functions or the dimensionality of the database on which the utility functions are defined. 

Next, we discuss how to measure the satisfaction of a user when shown a set of points rather than only one point.

\begin{definition}[Satisfaction and Best Point~\cite{Nanongkai:regret}]
Let $S$ be a subset of $D$
and $f$ be the utility function of a user.
$f$'s \textit{satisfaction} with respect to $S$ (or the satisfaction of the user with the utility function $f$ with respect to $S$), denoted by $sat(S, f)$,
is defined to be ${\max_{p\in S}f(p)}$.
$sat(S,f)$ is defined to $0$ if $S$ is empty.
A point $p$ is said to be $f$'s \textit {best point} in $S$ if $p = \arg\max_{p\in S}f(p)$.

\end{definition}

Consider a set $S$ containing ``Intercontinental'' and ``Hilton'' from Table~\ref{tab:example:hotelUtil}.
The utility of ``Hilton'' with respect to Alex is the greatest.
Thus, ``Hilton'' is Alex's best point in $S$.
Alex's satisfaction with respect to $S$ is equal to the utility
of ``Hilton'' with respect to Alex (i.e., 0.4).

When a set of a certain number of points is shown to a user,
the satisfaction of this user may decrease. In other words,
the dissatisfaction of this user may increase.

\begin{definition}[Regret and regret ratio~\cite{Nanongkai:regret}]
Let $S$ be a subset of $D$.
For a user whose utility function is $f$, when s/he sees the set $S$ instead of $D$,
the \emph{regret} of $f$ with respect to $S$, denoted by $r(S, f)$,
is defined to be $sat(D, f) - sat(S, f)$ and the \emph{regret ratio}
of $f$, denoted by $rr(S, f)$, to be ${\frac{r(S, f)}{sat(D, f)}}$.
\end{definition}

The regret ratio of a user with respect to a set $S$
captures how dissatisfied this user is if this subset $S$ of the database $D$, instead of the whole database, is shown to this user.

Let $F$ be the set of all possible utility functions.
There are two possible cases of $F$, the \emph{uncountable case of $F$} and
the \emph{countable case of $F$}. 
In the following, 
we focus on discussing the first case of $F$, and the second case is a simple extension of the discussion here and is analyzed in 
\ifx\techReport\undefined
our technical report \cite{techreport}.
\else
Section \ref{appendix:countable}. 
\fi

We let $\Theta$ denote the distribution of the utility functions in $F$, and let $\eta(f)$ be the probability distribution function for utility functions $f$ in $F$ corresponding to $\Theta$. Note that $\int_{f\in F} \eta (f)df = 1$.

Finding $\eta (f)$ is a typical machine learning problem widely explored in areas such as user's recommender systems \cite{recommender1, recommender2}, Bayesian learning models \cite{LearningGaussianProcesses, LearningGaussianProcesses2} and user's preference elicitations \cite{PreferenceElicitation1, PreferenceElicitation2}. For instance, Bayesian learning models discuss how the utility function of a user can be learned if the user has provided ratings on only a small subset of the database. In general, user ratings and other data such as users' activities recorded in logs can be used for this purpose, from which, we can build a statistical model and find $\eta(.)$. In Section {\ref{subsubsec:exp:experimentalResultRealDataset}}, we discuss how we tackled this problem in our experiments.

\begin{definition}[Average Regret Ratio]
Let $F$ be a set of users with the probability density function $\eta(.)$ corresponding to a probability distribution $\Theta$
and $S$ be a subset of $D$.
The \emph{average regret ratio} of $F$
for $S$, denoted by $arr(S)$, is defined to be $\int_{f\in F} rr(S, f)\eta(f)df$.
\end{definition}

In the above formulation, $arr(S)$ is the expected value of the regret ratio of a user when the users' utility functions follow the distribution $\Theta$. In the rest of this paper, we focus on the most general case of $F$ (i.e., when $F$ is the set of all continuous utility functions in the space $R^n\geq 0$) and we assume that the utility value for any point is at most 1. Note that the distribution $\Theta$ allows us to select for each instance of the problem which utility functions should be considered and how probable they are.

We are ready to present the problem discussed in this paper, called \emph{\underline{F}inding \underline{A}verage Regret \underline{M}inimizing Set (FAM)},
as follows.

\begin{problem}[\underline{F}inding \underline{A}verage Regret \underline{M}inimizing Set (FAM)]
Given a positive integer $k$, and a probability distribution $\Theta$
we want to find a set $S$ containing $k$ points in $D$ such that
$arr(S)$ is the smallest (i.e., $S = \arg\min_{S'\subseteq D, \left\vert{S'}\right\vert = k}arr(S')$).
\end{problem}

The following theorem shows that FAM is NP-hard for a general probability distribution.

\begin{theorem}
Problem FAM is NP-hard.
\label{thm:NPHardness}
\end{theorem}

\textit{Proof sketch.} By means of a reduction from Set Cover problem \cite{Karp1972} to FAM. 

\hfill\ensuremath{\square}

The NP-hardness result holds when the specification of the general probability distribution is allowed to be of non-constant size, that is, the probability distribution of FAM can be any general probability distribution and it does not have to be specified by at most a constant number of parameters.

Another relevant metric that can help us measure whether most of the regret ratio of the users are close to the average regret ratio or there are large variations from the mean is the variance of the regret ratios defined as follows.

\begin{definition}[Variance of Regret Ratio]
Let $F$ be a set of users with the probability density function $\eta(.)$
and $S$ be a subset of $D$.
The \emph{variance of regret ratio} of $F$
for $S$, denoted by $vrr(S)$, is defined to be $\int_{f\in F} (rr(S, f)-arr(S))^2\eta(f)df$.
\end{definition}

Although the FAM problem focuses on minimizing average regret ratio, it is important for a set $S$ to have low $vrr(S)$ as well. In our empirical studies (Section \ref{sec:exp}), we use estimates of $vrr(S)$ to compare different selection sets.

\subsection{Property Definition}\label{subsec:pro}

As we described before, we observe two interesting properties, namely
the ``supermodularity'' property  and the ``steepness'' property.
In this section, we define these properties.

The first property (i.e., the ``supermodularity'' property) is defined as follows.

\begin{definition}[Supermodularity~\cite{Vitorpi:supermodular}]
Let $U$ be a universe. A function $g: 2^U \rightarrow R_{\geq 0}$ is said to be \textit{supermodular} if and only if 
for any two sets, namely $S$ and $T$, where $S \subseteq T \subseteq U$, and 
for any element $x \in U - T$, we have $g(S \cup \{x\}) - g(S) \leq g(T \cup \{x\}) - g(T)$. A function $g(.)$ is said to be \textit{submodular} if and only if $-g(.)$ is supermodular.
\end{definition}

The second property (i.e., ``steepness'') requires the concept of ``monotonically decreasing function'' defined next.

\begin{definition}[Monotonically Decreasing (Set) Function~\cite{Vitorpi:supermodular}]
Let $U$ be a universe.
A function $g: 2^U \rightarrow R_{\geq 0}$ is said to be \textit{monotonically decreasing} if and only if  for any $A \subseteq U$ and each $x \in U - A$, $g(A \cup \{x\}) \leq g(A)$.
\end{definition}

We are ready to define the ``steepness'' property as follows.

\textit{Steepness}, as defined by \cite{Vitorpi:supermodular}, is the maximum possible marginal decrease of the function. 

\begin{definition}[Steepness~\cite{Vitorpi:supermodular}]
Let $U$ be a universe and $g$ is a function $2^U \rightarrow R_{\geq 0}$.
For any set $X \subseteq U$ and an $x \in X$,
we define a function $d(x, X) = g(X - \{x\}) - g(X)$.
The \textit{steepness}, $s$, of function $f$ is defined to be $\max_{x \in U, d(x, \{x\}) > 0} \frac{d(x, \{x\}) - d(x, U)}{d(x, \{x\})}$.
\end{definition}

\section{Algorithm for General Case}
\label{sec:algo}
In this section, we focus on the general case of FAM, when the set of utility functions is continuous and can have any probability distribution. In this case, the problem is NP-Hard and is unlikely to have a polynomial-time algorithm that solves the problem optimally. Thus, we focus on providing an approximate algorithm for the problem. We will first give the ``supermodularity'' property used in algorithm {\scshape{Greedy-Shrink}}
in Section~\ref{subsec:sup} and then present algorithm {\scshape{Greedy-Shrink}} in Section~\ref{algo:greedy}.
Finally, we give the detailed steps of  algorithm {\scshape{Greedy-Shrink}} in Section~\ref{subsec:alg:detailedStep}.

\subsection{Supermodularity and Monotonically Decreasing}\label{subsec:sup}
We first give the ``supermodularity'' property for our problem as follows.
\begin{theorem} 
$arr(\cdot)$ is a supermodular function.
\label{thm:supermodularity}
\end{theorem}

\textit{Proof sketch.} We need to show that for all $S \subseteq T \subseteq D$ and for any element $p\in D - T$, $arr(S \cup \{p\}) - arr(S) \leq arr(T \cup \{p\}) - arr(T)$. To do so, consider an element $p \in D - T$. There are two possibilities depending on whether $p$ is the best point in $S$ for any user or not. If $p$ is not the best point in $S\cup\{p\}$ (and consequently, since $S\subseteq T$, not the best point in $T\cup\{p\}$ either) for any utility function, then $arr(S \cup \{p\}) - arr(S)$ and $arr(T \cup \{p\}) - arr(T)$ are both zero which proves the result in this case. 

Otherwise, if $p$ is the best point in $S\cup\{p\}$ for some utility functions, then, using the definition of regret ratio we can write $arr(S \cup \{p\}) - arr(S)=-\int_{f \in U}\frac{sat(S \cup \{p\}, f) - sat(S, f)}{sat(D, f)}\eta (f)df$ where $U$ is the set of utility functions whose best point changes when $p$ is added to $S$. Similarly, $arr(T \cup \{p\}) - arr(T)=-\int_{f \in U}\frac{sat(T \cup \{p\}, f) - sat(T, f)}{sat(T, f)}\eta (f)df$ holds because of the same reasoning and that if the best point of a user changes when $p$ is added to $T$, the user must be in $U$ (since $S$ is a subset of $T$). We can show that $\int_{f \in U}\frac{sat(S \cup \{p\}, f) - sat(S, f)}{sat(D, f)}\eta (f)df$ is more than or equal to $\int_{f \in U}\frac{sat(T \cup \{p\}, f) - sat(T, f)}{sat(T, f)}\eta (f)df$ because $S$ is a subset of $T$ which implies that $arr()$ is a supermodular function.

\hfill\ensuremath{\square}

Intuitively, if we have a set $S$ of points and add one point to $S$, 
there is a higher chance of increasing the satisfaction 
of a user on average and thus decreasing the regret ratio of this user on average (i.e., the average regret ratio for this set including the additional point) 
compared with the case when we have a larger set $T$ (i.e., $S \subseteq T$) and add the same point to $T$.

It is easy to verify that the average regret ratio (i.e., $arr(\cdot)$) is a monotonically decreasing function, as shown in the following lemma,
because adding a new point into a set either could reduce or could not change the average regret ratio for this set.

\begin{lemma} 
$arr(\cdot)$ is a monotonically decreasing function.
\label{thm:monotonicallyNonIncreasingFunction}
\end{lemma}

\textit{Proof sketch.} Similar to Theorem \ref{thm:supermodularity}, by dividing the problem into two cases depending on whether the newly added point is the best point for any user or not.

\hfill\ensuremath{\square}

\subsection{Algorithm {\scshape{Greedy-Shrink}}}\label{algo:greedy}
In this section, we present algorithm {\scshape{Greedy-Shrink}}. Algorithm {\scshape{Greedy-Shrink}} initializes
the solution set $S$ to the whole database and iteratively removes one point from the current solution set $S$ so that the average regret ratio of the resulting set is the smallest. This continues until the number of remaining points in $S$ is at most $k$. Pseudo-code of algorithm {\scshape{Greedy-Shrink}} is shown in Algorithm~\ref{algo}.

\begin{algorithm}[tb]
\caption{Algorithm {\scshape {Greedy-Shrink}}} \label{algo}
\begin{algorithmic} [1]
\STATE $S \leftarrow D$\
\WHILE{$|S| > k$}
	\STATE $p' \leftarrow \arg\min_{p \in S} arr(S - \{p\})$\;
	\STATE $S \leftarrow S - \{p'\}$\;
\ENDWHILE
\STATE \textbf{return} $S$
\end{algorithmic}
\end{algorithm}

We use the second interesting property called the ``steepness'' property as follows.
This property gives us some ideas about the approximate ratio of algorithm {\scshape{Greedy-Shrink}}.

\begin{theorem}
Let $S$ be the set returned by {\scshape{Greedy-Shrink}} (i.e., Algorithm~$\ref{algo}$) and $S_o$ be the optimal solution of problem FAM.
The approximate ratio of {\scshape{Greedy-Shrink}} (i.e., $\frac{arr(S)}{arr(S_o)}$) is
$\frac{e^t - 1}{t}$, where $t = \frac{s}{1-s}$ and $s$ is the steepness of $arr(\cdot)$.
\label{thm:steepness}
\end{theorem}

\textit{Proof.} Based on \cite{Vitorpi:supermodular}, minimizing a monotonically decreasing supermodular function with the steepness $s$, using an algorithm that at each iteration removes a point whose removal increases the value of the function the least, will result in a solution which is within $\frac{e^t - 1}{t}$ factor of the optimal solution, where $t = \frac{s}{1-s}$. Then, the result follows from Theorem ~\ref{thm:supermodularity} and Lemma ~\ref{thm:monotonicallyNonIncreasingFunction}.

\hfill\ensuremath{\square}

Although it is shown that there is an approximate ratio (which could be greater than 1),
in our experiments on small datasets, the \emph{empirical} approximate ratio of
{\scshape{Greedy-Shrink}} is exactly 1. This could be possibly explained
by the loose theoretical bound of the approximate ratio in Theorem~\ref{thm:steepness}.

\subsection{Detailed Steps}
\label{subsec:alg:detailedStep}

\label{sec:sample}

In this section, we describe how to compute the average regret ratio
for a given set $S$ containing points in $D$.
There are two challenges.
The first challenge is the uncountable space of $F$ (i.e., a set of all possible
utility functions) used to evaluate the average regret ratio.
The second challenge is the efficiency issue of evaluating the average regret ratio.

\smallskip
\noindent
\textbf{Challenge 1 (Uncountable Space of $F$):}
Note that $F$ is the set of all possible utility functions and is uncountable.
Therefore, we need to compute an integral when computing the average regret ratio
for a given set. Besides the complexity of such a computation, the solution of the integral relies on the distribution $\Theta$ of $F$, which means that solving the integral in the definition of the average regret ratio will be dependent on the choice of $\Theta$. However, in this section, we aim at providing a solution that can be applied to any probability distribution and for any choice of $\Theta$.

In this paper, we present a \emph{sampling} technique to compute the average regret ratio
with a theoretical bound.
Specifically, we need to determine the sampling size $N$ denoting the total number
of utility functions in $F$ to be sampled according to distribution $\Theta$.
The exact formula of computing $N$ is shown later.
With these $N$ sampled utility functions, we compute the estimated average regret ratio
by averaging the regret ratio of the sampled utility functions.

This estimated average regret ratio is similar to the exact average regret ratio
if the sampling size $N$ is determined carefully.
Let $\epsilon$ and $\sigma$ be an error and confidence parameter $\in [0, 1]$. We have the following theorem about the theoretical bound on the estimated average regret ratio.

\begin{theorem}
Let $arr$ be the estimated average regret ratio and $arr^*$ be the exact average regret ratio.
If $N \ge \frac{3\ln(\frac{1}{\sigma})}{\epsilon^2}$, then with the confidence at least $1-\sigma$,
\begin{equation*}
\left\vert arr - arr^* \right\vert < \epsilon
\end{equation*}
\label{thm:samplingTheory}
\end{theorem}

\textit{Proof sketch.} We use Chernoff bounds to prove the theorem. Let $X_1$ to $X_N$ be independent and identically (according to $\Theta$) distributed random variables denoting the regret ratio of a utility function, and let $X = \sum_{i=1}^NX_i$. By Chernoff bounds we have $Pr[\,X - E[X] \geq \epsilon'E[X]]\, \leq e^{\frac{-\epsilon'^2}{3}E[X]}$ for a constant $\epsilon'$. By substitution, we can obtain $Pr[\,X - E[X] \geq \sqrt{3E[X] \ln \frac{1}{\sigma}}]\,  \leq \sigma$ for a constant $\sigma$. Note that by definition and using linearity of expectation, $E[X] = N\times arr^*$ and $\frac{1}{N}\sum_{i=1}^N X_i = arr$. Therefore, substituting $arr^*$ and $arr$, and since $arr^* \leq 1$, we can obtain $Pr[\, arr - arr^* \geq \sqrt{\frac{3\ln \frac{1}{\sigma}}{N}}]\,  \leq \sigma$, which proves the theorem by letting $\epsilon=\sqrt{\frac{3\ln \frac{1}{\sigma}}{N}}$. 

\hfill\ensuremath{\square}

Let $F_N$ be the set of all $N$ sampled utility functions in $F$, sampled according to the distribution $\Theta$.
Thus, given a solution set $S$, we compute $arr(S)$ with the following equation.

\begin{equation}
arr(S) = \frac{1}{N} \sum_{f\in F_N}  \frac{\max_{p\in D}f(p) - \max_{p\in S}f(p)}{\max_{p\in D}f(p)}
\label{eqn:samplingFormula}
\end{equation}

\smallskip
\noindent
\textbf{Challenge 2 (Efficiency Issue):}
The second challenge is improving the efficiency of computing the average regret ratio. We employ two methods to improve the efficiency of computing average regret ration in practice. Firstly, at each iteration of the algorithm, when calculating average regret ratio, we only compute the best point for utility functions whose best point changes in the solution set. Secondly, we use the average regret ratio calculation at a previous iteration to prune the set of points considered at each iteration of the algorithm. The details of these two methods are discussed in \ifx\techReport\undefined
our technical report \cite{techreport}.
\else
Section \ref{appendix:improve}. 
\fi

\subsection{Theoretical Analysis}

In this section, we give some theoretical analysis of
algorithm {\scshape{Greedy-Shrink}}.

\subsubsection{Theoretical Bound}
Firstly, we give the theoretical guarantee on the solution set returned
by algorithm {\scshape{Greedy-Shrink}}.

\begin{theorem} Let $D$ be a set of points, $F$ a set of the utility functions with the probability distribution $\Theta$ and $k$ the number of desired representative points from the database.
The sampling of the utility functions based on the error and confidence parameters $\epsilon$ and $\sigma$ and running Algorithm $\ref{algo}$ results in the solution set  $S$, for which $arr^*(S) < \frac{e^t - 1}{t}(arr^*(S_O) + \epsilon) + \epsilon$ with the confidence of 1 - $\sigma$, where $arr^*(S)$ is the true value of the average regret ratio (not the value calculated by sampling).
\label{thm:discreteDistribution}
\end{theorem}

\subsubsection{Time Complexity}
Secondly, we analyze the time complexity of algorithm {\scshape{Greedy-Shrink}}.
We divide the time complexity of the algorithm in to two sections \textit{preprocessing time} and \textit{query time}. The preprocessing section corresponds to the steps required bofore the algorithm can be run. For instance, it consists of building any indexing data structure needed such as finding the best points in $D$ for all the users. The query time corresponds to the time it takes for the algorithm to run after all the data structures are created.

\textit{Preprocessing time.} In the preprocessing step, we need to sample $N$ utility functions which takes time $O(nN)$. Any data structure used for faster data retrieval from the database need to be created at this point as well. For each user, we need to find his/her best point in $D$ and for each point $p$, we also keep track of the users whose best point is $p$. We need to check all the $n$ points in the database of size $n$ to find the best point of a user. Thus, finding the best point for all the users takes $O(Nn)$. Therefore, the preprocessing step of the algorithm takes time $O(Nn)$.

\textit{Query time.} Algorithm $\ref{algo}$ has $n - k$ iterations and at each iteration $i$, $0 \leq i < n - k$, we need to calculate the average regret ratio $n - i$ times at the worst case. For each calculation of the average regret ratio, we need to find the best point in $S$ for each user. For each user, it will be required to go through a total of $n - i - 1$ points. Thus, the running time of the algorithm is $O(\sum_{i = 0}^{n - k - 1} (n - i)N(n - i -1))$.  Hence, the worst case running time of the algorithm is $O(\sum_{i = 0}^{n - k - 1} N(n - i)(n - i - 1))$ = $O(Nn^3)$. So, in the worst case, the total running time of the preprocessing and query steps of the algorithm combined is $O(Nn^3)$. Note that, there are heuristics put in practice to reduce the running time on average explained in Section \ref{subsec:alg:detailedStep}.

\subsubsection{Space Complexity}
Thirdly, we analyze the space complexity of algorithm {\scshape{Greedy-Shrink}}.
If we are given the utility scores for each user, we will need $O(nN)$ space to store the data. We also need to store the solution set, which takes space $O(n)$ because the solution set is initially the same as the database and is eventually reduced to size $O(k)$. Moreover, for each point $p$ in the database, we store the users whose best point is $p$. For this, we use a linked list of users for each point, that is, an array of linked lists. The length of the array will be $O(n)$, because for each point an entry will be required in the array. Since there is exactly one best point for each user (that we keep track of), the total length of all the linked lists will be $O(N)$. Thus, the total space required to store these information is $O(nN)$.

Note that if for a $d$-dimensional database we are given the utility functions of users in a form (for instance linear utility functions) that can be stored in $O(dN)$, then we can reduce the space requirement to $O(d(N + n))$, as the database itself can be stored in $O(dn)$. Note that this will increase the time complexity of the algorithm by a factor of $d$ as calculating the utility score for each point will now require $O(d)$ time.
\section{Algorithm on Dataset Containing Two Dimensions}
\label{sec:twoDimAlgo}
So far we have shown that the FAM problem is NP-hard for a general continuous probability distribution and provided an approximation algorithm for such a general case. However, special cases of the problem are of interest both in practice and in theory. 
Here, we consider a special case of the problem with continuous distribution of linear utility functions and provide an exact algorithm that can solve the FAM problem optimally when the dimensionality of the database is two. Two dimensional databases can arise in practice when there are only two features available for the data, or after feature selection or extraction from a larger set of features.

\subsection{Linear Utility Functions}\label{sec:dyn:linearUtility}
First, note that the linear utility functions are of the form $f(p) = w_1p[1] + w_2p[2]$ where $p[1]$ and $p[2]$ are the first and second attributes of the point $p$, and $w_1$ and $w_2$ are the weights of the utility function for each dimension. We can consider $(w_1, w_2)$ as a vector, and it is easy to see that scaling the vector does not change the regret ratio of a utility function from any set. Hence, we only need to consider the direction of the vector, which we can measure by the angle it makes with the first dimension. Therefore, in this section, an angle $\theta$ is used to represent the set of utility functions that make the angle $\theta$ with the first dimension, that is, $\theta = \arctan(\frac{w_2}{w1})$. We let $F_{\theta_l}^{\theta_u}$ be the set of utility functions whose angle is between ${\theta_l}$ and ${\theta_u}$. For ease of notation, we define, for any $\theta$, $F_\theta = F_\theta^{\frac{\pi}{2}}$.

In our discussion in this section, we make sure our dataset only includes skyline points and that the points are sorted in descending order of their first dimension. Therefore, if $i<j$, then $p_j[1] \leq p_i[1]$ and because they are in skyline, $p_j[2] \geq p_i[2]$. Moreover, we limit the set of utility functions where $0\leq w_1, w_2\leq 1$.

Now, given two points $p_i, p_j \in D$, $i > j$, to find utility functions that prefer $p_i$ over $p_j$, we need to solve $w_1p_i[1] + w_2p_i[2]>w_1p_j[1] + w_2p_j[2]$, which gives us $\frac{w_2}{w_1}>\frac{p_j[1] - p_i[1]}{p_i[2] - p_j[2]}$, and changing the direction of the inequality, we can find an expression for the utility functions that prefer $p_j$ over $p_i$. Let $\theta_{i, j} = \arctan(\frac{p_j[2] - p_i[2]}{p_i[1] - p_j[1]})$. Then, consider a utility function, $f$ with angle $\theta$. If $\theta>\theta_{i, j}$, then $f(p_i)> f(p_j)$; if $\theta<\theta_{i, j}$, then $f(p_i)<f(p_j)$; and if $\theta=\theta_{i, j}$, then $f(p_i)=f(p_j)$. This means that to see whether a utility function prefers a point $p_i$ over $p_j$ or not, we only need to compare its angle with $\theta_{i, j}$. $\theta_{i, j}$ divides the space of utility functions into two subspaces based on whether they prefer $p_i$ over $p_j$ or not and $\theta=\theta_{i, j}$ is the equation of a line on the $(w_1,w_2)$-plane that passes through origin. Note that if $i < j$, the direction of all the inequalities would be reversed. Finally, we let $\theta_{i, n+1} = \frac{\pi}{2}$ for simplicity of notation.

\subsection{Recursive Formulation}\label{sec:dyn:recurrence}

Let $arr^*(r, i, \theta)$ be the optimal solution to the following problem: given that the point $p_i$ is already selected and is the best point for utility function $\theta$, choose at most $r$ points to minimize the average regret ratio of users in $F_{\theta}$. It is easy to see that the optimal solution to FAM can now be written as $\min_{1\leq i\leq n} arr^*(k-1, i, 0)$, because one of the points in the solution has to be the best point for $\theta=0$ and we are checking all possible points and selecting the minimum. 

Moreover, let $arr(S, F_{\theta_l}^{\theta_u})$ be the average regret ratio of the set $S$ over the utility functions in $F_{\theta_l}^{\theta_u}$.

Intuitively, to solve the problem of $arr^*(r, i, F_\theta)$, we can first find the subset of $F_{\theta}$ for which $p_i$ is the best point in the optimal solution. Then, for such a subspace, the average regret ratio only depends on $\{p_i\}$ and none of the other points in the optimal solution. Then, for the rest of the utility functions, we can solve the problem recursively. Next, we provide a recurrence relation to solve the problem. 

\begin{theorem}
        Given an integer $r$, and an angle $\theta_l$, $0\leq\theta_l\leq\frac{\pi}{2}$. Then, with base cases $arr^*(0, i, \theta_l) = arr(\{p_i\}, F_{\theta_l}^{\frac{\pi}{2}})$ and $arr^*(r, i, \frac{\pi}{2}) = 0$, it hods that
$arr^*(r, i, \theta_l) = \min_{i<j\leq n+1, \theta_{i, j} \geq \theta_l} arr(\{p_i\}, F_{\theta_l}^{\theta_{i, j}}) + arr^*(r-1, j, \theta_{i, j})$
\end{theorem}

\textit{Proof.} First note that because $p_i$ is the best point for $\theta_l$, in the optimal solution, it must be the best point for a range of utility functions $F_{\theta_l}^\theta$ for some value $\theta$. Let $p_j$ be the point in the optimal solution such that it is the best point for the utility functions in the range $F_{\theta}^{\theta'}$ for some value $\theta'$. Then $\theta$ has to be equal to $\theta_{i, j}$, as discussed in Section \ref{sec:dyn:linearUtility}, because it is the angle separating utility functions who prefer $p_i$ over $p_j$. Note that if such a $p_j$ does not exist, it means that either $\theta = \frac{\pi}{2}$ (because $p_i$ has to be the best point for all the utility functions), for which we can let $j=n+1$, and then the optimal solution will be $arr(\{p_i\}, F_{\theta_l}^{\frac{\pi}{2}})$, or that $r$ is equal to zero. For the latter case, we use the base case where $arr^*(0, i, \theta_l) = arr(\{p_i\}, F_{\theta_l}^{\frac{\pi}{2}})$.

Now, because $p_i$ is the best point for the utility functions in $F_{\theta_l}^{\theta_{i, j}}$, the average regret ratio for users in $F_{\theta_l}^{\theta_{i, j}}$ is $arr(\{p_i\}, F_{\theta_l}^{\theta_{i, j}})$. For the users in $F_{\theta_{i, j}}$, we know that $p_j$ is in the optimal solution, and is the best point for utility functions $\theta_{i, j}$. Therefore, for users in $F_{\theta_{i, j}}$, the optimal solution is $arr^*(r-1, j, F_{\theta_{i, j}})$.

Finally, we do not know the value of $j$, but it has to satisfy the following properties. First, $\theta_{i, j} \geq \theta_l$ because $p_i$ has to be the best point for $\theta_l$. Secondly, it must be the case that $i<j$, because $p_j$ has to be the best point over a range of utility functions with angles larger than $\theta_{i, j}$ (if $j<i$, $p_j$ will be the best point over a range of utility functions with angles smaller than $\theta_{i, j}$, as discussed in Section \ref{sec:dyn:linearUtility}). Therefore, $j$ has to be one of the values in the range $i<j\leq n+1, \theta_{i, j} \geq \theta_l$. Thus, we can go through all the at most $n$ possible values and choose the one with minimum value. This proves the recurrence relation. 

\hfill\ensuremath{\square}

\subsection{Dynamic Programming Algorithm}
\subsubsection{Overall Algorithm}\label{sec:dyn:overallAlgo}
We use the recurrence relation in Section \ref{sec:dyn:recurrence}, to solve the problem. First we find the skyline of the dataset and sort the points by their first dimension. 

Consider calculating $arr(\{p_i\}, F_{\theta_l}^{\theta_{i, j}})$. Note that $\theta_l$, in the recursive call is always equal to $\theta_l= \theta_{i, z}$ for some value of $z$, except for the first function call when $\theta_l = 0$. Therefore, for every $i$, there are at most $n+1$ possible values for $\theta_l$ and there are at most $n+1$ possible values for ${\theta_{i, j}}$ (because $j$ can be equal to $n+1$). This means that we can precompute all the values and store them. We discuss how the values are actually calculated in Section \ref{sec:twoDimAlgo:calcARR}.

Note that calculating $\theta_{i, j}$, as described in Section \ref{sec:dyn:linearUtility} takes constant time so we can calculate them on the go.

After computing $arr(\{p_i\}, F_{\theta_l}^{\theta_{i, j}})$ for all $i$, $\theta_l$ and $\theta_{i, j}$ values, then, for all $i$, we start with $arr^*(k-1, i, 0)$ and recursively solve the problem, storing $arr^*(k-1, i, 0)$ in an array so that each element of the array is calculated only once. Finally, when for all $i$, $arr^*(k-1, i, 0)$ has been calculated, we go through all the $n$ possible values and choose the one with the smallest average regret ratio as the optimal solution.

\subsubsection{Calculation of Average Regret Ratio}\label{sec:twoDimAlgo:calcARR}
We also need to address the issue of calculating $arr(\{p\}, F_{\theta_l}^{\theta_{u}})$ for different $p$, $\theta_l$ and $\theta_u$ values. For this, by definition, we get $arr(\{p\}, F_{\theta_l}^{\theta_{u}}) = \int_{f\in F_{\theta_l}^{\theta_{u}}}(1-\frac{f(p)}{\max_{p'\in D}f(p')})\eta(f)df$. The issue in solving the integral is the $\max$ function in the denominator that complicates the integral. Intuitively, to address this issue, we divide the integral into at most $n$ regions depending on which point is a utility function's best point in $D$. Then, we can split the integral into sum of $n$ integral, for each of which we know the point, $p_i$, that satisfies $f(p_i) = \max_{p'\in D}f(p')$, and hence we can replace the $\max_{p'\in D}f(p')$ function by $f(p_i)$. 

Let $\theta_i^l = \max_{j<i\leq n}\theta_{i, j}$ and let $\theta_i^u=\min_{i<j\leq n}\theta_{i, j}$. Then, as discussed in Section \ref{sec:dyn:linearUtility}, a utility function with angle $\theta$, $\theta > \theta_i^l$ will prefer $p_i$ over all points $p_j$, $j<i$. Moreover, a utility function with angle $\theta$, $\theta < \theta_i^u$ will prefer $p_i$ over all points $p_j$, $j>i$. Thus, $p_i$ is the best point for all the utility functions with angles in the range $[\theta_i^l, \theta_i^u]$, or the set $F_{\theta_i^l}^{\theta_i^u}$. 

Moreover, let $t_i^l = \max\{\theta_i^l, \theta_l\}$ and $t_i^u = \min\{\theta_i^u, \theta_u\}$. Note that the range $[t_i^l, t_i^u]$ is the set of utility functions in $F_{\theta_l}^{\theta_{u}}$ whose best point is $p_i$. We can write $F_{\theta_l}^{\theta_{u}} = \bigcup_{1\leq i\leq n}F_{t_i^l}^{t_i^u}$ (we can simply ignore the cases where $t_i^l>t_i^u$). 

Finally, let $c_i^l = \tan(t_i^l)$ and $c_i^u = \tan(t_i^l)$. $c_i^l$ and $c_i^u$ define two half plane such that for a utility function $(w_1, w_2)$ to be in $[t_i^l, t_i^u]$, it has to satisfy $w_2\geq c_i^l w_1$ and $w_2\leq c_i^u w_1$. As a result of this, we can write $arr(\{p\}, F_{\theta_l}^{\theta_{u}})$ as 

$$
\sum_{1\leq i\leq n}\int_0^1\int_{c_i^lw_1}^{c_i^uw_1}(1-\frac{(w_1p[1]+w_2p[2])}{w_1p_i[1]+w_2p_i[2]})\eta(f)dw_2dw_1
$$

In the above equation, we have used $F_{\theta_l}^{\theta_{u}} = \bigcup_{1\leq i\leq n}F_{t_i^l}^{t_i^u}$ to split the integral into $n$ regions. We can do this because the overlapping region for each $F_{t_i^l}^{t_i^u}$ and $F_{t_j^l}^{t_j^u}$ corresponds to the equation of a line, which has no effect on the integration. We have also imposed the limits on the utility functions that  $0\leq w_1, w_2\leq 1$, but we have assumed that $c_i^u \leq 1$. If this is not the case, we need to subtract the extra region for which $w_2>1$ from the integration, that can be done in a similar manner. Here, we only focus on $c_i^u \leq 1$.

The exact calculation of the average regret ratio also depends on the choice of $\eta(f)$. For instance, for a uniform distribution where $\eta(f) = 1$, then we can integrate the expression exactly and provide a closed-form solution for each integral. We do not provide detail of the integration for this case as the solution does not have a compact form, but there are constant number of terms in the closed from solution, resulting in the evaluation of each integral in constant time (note that there are a total of $n$ integrals to be computed for the calculation of the average regret ratio). However, the integral does not necessarily have a closed-form solution for different choices of $\eta(f)$. Therefore, sampling methods as discussed in Section \ref{sec:sample} might still be useful for this case. 

\subsubsection{Time Complexity}
First, the algorithm finds the skyline points, sorts them and computes $arr(\{p_i\}, F_{\theta_l}^{\theta_{i, j}})$ for all $i, j, \theta_l$, which takes $O(n^4)$ as there are total of $O(n^2)$ different possibilities for $\theta_l$ and $j$ together, as discussed in Section \ref{sec:dyn:overallAlgo}. Filling the $arr^*$ table requires filling $O(kn^2)$ elements, each of which take $O(n)$, which is $O(kn^3)$. Finally, finding an $i$ for which $arr^*(k-1, i, 0)$ is minimum needs a linear scan of the elements and $O(n)$ time. Therefore, overall, the algorithm takes $O(n^4)$.

\begin{table*}
\small
\begin{tabular}{c c c}
    \hspace*{-1cm}
    \begin{minipage}[t]{0.4525\textwidth}
		\centering\begin{tabular}[t]{| c | c | c | } \hline
		$S_{arr}$  &  $S_{mrr}$ & $S_{k-hit}$ \\ \hline
		Stephen Curry & LaMarcu Aldridge & Stephen Curry  \\
		Kevin Durant  & DeMarcus Cousins & Kevin Durant \\ 
		James Harden & Stephen Curry & James Harden \\ 
		DeAndre Jordan &George Hill &  Draymond Green \\ 
		Russell Westbrook & Ramon Sessions & Russell Westbrook  \\ \hline 
		\end{tabular}
        \vspace*{0.1cm}
         \caption{Three Sets of 5 players computed based on the average regret ratio (arr), the maximum regret ratio (mrr) and the $k$-hit query ($k$-hit) (i.e., $S_{arr}, S_{mrr}$ and $S_{k-hit}$)}
		\label{table:6}
		\label{tab:differentSetResult}		
	\end{minipage}
	\hfill
 &
    \hspace*{0.3cm}
    \begin{minipage}[t]{0.29\textwidth}
		\centering\begin{tabular}[t]{| c | c |} \hline
			Top 1 to 5 & Top 6 to 10\\\hline
			Stephen Curry & Derick Rose\\
			LeBron James & Russell Westbrook \\
			Kobe Bryant & Kyrie Irving \\
			Kristaps Porzingis & James Harden \\
			Kevin Durant & Jimmy Butler \\\hline
		\end{tabular}
        \vspace*{0.1cm}
        \caption{Top 10 NBA players in 2016 according to the number of jerseys sold}
		\label{table:Popular}
	\end{minipage}	
	\hfill
 &
    \hspace*{0.2cm}
	\begin{minipage}[t]{0.2\textwidth}
		\centering\begin{tabular}[t]{|c | c | c |} \hline
			Dataset              &	$d$    & $n$ \\ \hline
			Household-6d     &	6    & 127,931   \\\hline
			Forest Cover      &11   & 100,000        \\\hline
			US Census          &	10  & 100,000        \\\hline
			NBA                    &15   & 16,915	      \\\hline
			Yahoo!Music       &	-    & 8,933        \\\hline
		\end{tabular}
        \vspace*{0.1cm}
		\caption{Real datasets' information}
		\label{table:3}
		\label{tab:realDatasetStat}
	\end{minipage}	
\end{tabular}
\vspace*{-0.8cm}
\end{table*}
\section{Empirical studies}\label{sec:exp}

We conducted experiments on a workstation with 2.26GHz CPU and 32GB RAM. All programs were implemented in C++.
The default value of the sampling size, $N$, for evaluating
the average regret ratio of a given set is set to 10,000.

In Section~\ref{subsec:exp:arrVSmrr}, we first compare the solution set based on the average regret ratio studied in this paper
with two solution sets studied in previous papers,
namely the solution set based on the maximum regret ratio~\cite{Nanongkai:regret} and
the solution set based on the $k$-Hit query~\cite{wong:kHit}, to compare the usefulness
of the solution set discussed in this paper compared with existing studies.
Then, we present the experimental results based on the average
regret ratio in Section~\ref{subsec:exp:arrResult}.

\subsection{Avg. Regret Ratio vs. Max. Regret Ratio vs. k-Hit}
\label{subsec:exp:arrVSmrr}
In this experiment,  we used the NBA dataset from 2013 to 2016.
In this dataset, there are 22 dimensions about the statistical records
of NBA players (including the number of points scored and the number of blocks).
There are totally 664 players.
In this experiment, the utility functions used are linear
and since we did not have access to any information that could help us model the distribution of the utility functions, we set it to be a uniform distribution.

According to this dataset, we executed our proposed algorithm designed for the average regret ratio
to generate a set $S_{arr}$ of 5 players as the result based on the average regret ratio.
We also executed the algorithm \cite{Nanongkai:regret} designed for the maximum regret ratio
to generate a set $S_{mrr}$ of 5 players as the result based on the maximum regret ratio.
Furthermore, we executed the $k$-hit algorithm \cite{wong:kHit}
to generate a set $S_{k-hit}$ of 5 players as the result based on the $k$-hit query.
These three sets could be found in Table~\ref{tab:differentSetResult}.

In this experiment, we compare the ``goodness'' of the set $S_{arr}$
with the two sets $S_{mrr}$ and $S_{k-hit}$
based on not only
an online survey manner (which could be regarded as ``subjective'')
but also an external statistics manner (which could be regarded as ``objective'').

Firstly, we conducted an online survey whose setup is
similar to \cite{UserStudy}. We set up an online survey in ``Amazon Mechanical Turk'' to ask
participants with basic NBA knowledge for their favorite NBA players where we paid each participant $\$0.05$.
There are totally 890 participants in this survey.

In this survey, there are 2 questions.
The purpose of the first question is to filter out
all participants without basic NBA knowledge from participating in
the survey. This question is a simple basketball question which shows the picture
of ``Stephen Curry'', a famous NBA player, and asks each participant who the player is.
After removing all participants who answered this question wrongly,
there are 702 participants 
and we regard them as the participants with basic NBA knowledge
and their responses are used in our experimental results.
The second
question is to show
three sets, namely $S_{arr}$, $S_{mrr}$ and $S_{k-hit}$, and to ask
the participants to select one set which collectively contains
better players in their opinion.
This question can be used to determine whether $S_{arr}$ is better than $S_{mrr}$ and $S_{k-hit}$.

Secondly, we also compare the ``goodness'' of the set $S_{arr}$
with both $S_{mrr}$ and $S_{k-hit}$
based on the
external statistics about NBA player jersey sales in 2016~\cite{NBA-News}.
Table~\ref{table:Popular} shows the top-5 and the top-10 NBA players
according the number of jerseys sold in 2016. Note that the jersey sale should not be regarded as the ``ground truth" of our problem. Instead, similar to other modelling problems \cite{QRH17, YHT16, RSG16} in the data mining and the information retrieval 
community, this information could be regarded as a reference source of information on whether what we found is useful or practical for the real world.

Consider the survey result first.
According to the response to the second question in the survey, about 56\%, 17\% and 27\% of the participants preferred $S_{arr}$, $S_{mrr}$ and $S_{k-hit}$, respectively, which
suggests that the result based on the average regret ratio is more preferred compared
with the other two results. 
Note that $S_{arr}$ and $S_{k-hit}$ contain the same 4 players except one where 
``DeAndre Jordan'' is in $S_{arr}$ and ``Draymond Green'' is in $S_{k-hit}$.
Although these two sets differ only in one player,
there is almost 30 percent difference between $S_{arr}$ and $S_{k-hit}$ which could be possibly explained
with the following two reasons. 
Firstly, ``Stephen Curry'' and ``Kevin Durant'' (which are in both sets) play in the same team as ``Draymond Green'', which makes set $S_{k-hit}$ 
less representative, as opposed to including ``DeAndre Jordan'' who plays in a different team. 
Secondly, DeAndre Jordan's position as a center and his better performance in rebounds (he has been among the top-3 players with the most rebounds in the past 3 seasons) complements the capabilities of the other 4 selected players in $S_{arr}$, while Draymond Green's position as a power forward and his performance in different statistics overlaps with the other players in $S_{k-hit}$. 

Next, consider the result based on the external statistics about NBA player jersey sales (Table~\ref{table:Popular}).
Surprisingly, 4 players out of 5 players in $S_{arr}$ and $S_{k-hit}$ are in the top-10 players based on the number of jerseys
sold (Table~\ref{table:Popular}).
They are Stephen Curry, Kevin Durant, Russell Westbrook and James Harden.
Besides, 2 players out of 5 players in $S_{arr}$ and $S_{k-hit}$ are in the top-5.
They are Stephen Curry and Kevin Durant.
However, only 1 player out of 5 players in $S_{mrr}$ is in the top-5 and the top-10.
He is Stephen Curry.

Finally, note that in our selection set $S_{arr}$, DeAndre Jordan (not in the list of top-10 jersey sales) plays in the center position and does not score many points, but has a very high number of rebounds (higher than any of the players in the top-10 list). The other 4 players (who were in the list of top-10 jersey sales) had high scoring performance but played in different positions such as point guard, small forward and shooting guard (furthermore, they, together, had a diverse set of statistics that they performed well in). As a result, the set can be regarded as a representative of NBA players and can satisfy the expectation of different NBA fans who pay attention to different statistics.

\begin{figure}[t]
	\includegraphics[width=\columnwidth]{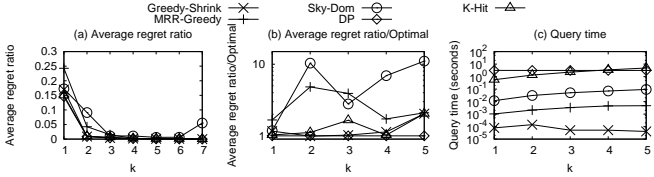}
	\caption{Effect of $k$ on 2-dimensional dataset}
	\label{fig:DP:k}
	\vspace*{-0.4cm}
\end{figure}

\begin{figure}[t] 
  \centering
  \includegraphics[width=0.75\columnwidth]{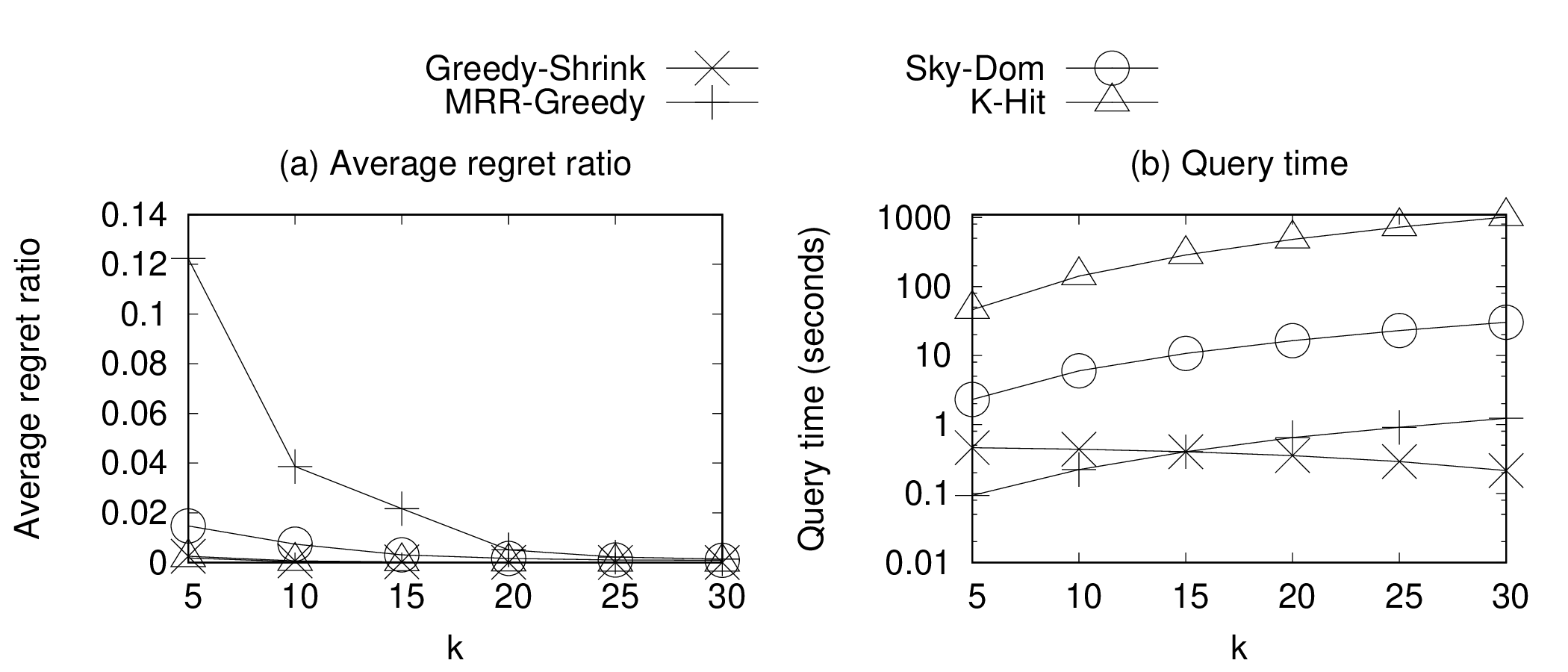}
  \caption{Effect of $k$ on Yahoo! Dataset}
  \label{fig:yahooDataset:k}
    \vspace{-0.4cm}

\end{figure}
\begin{figure}[t]
\centering
\begin{tabular}{c c}
\hfill
  \includegraphics[width=0.4\columnwidth]{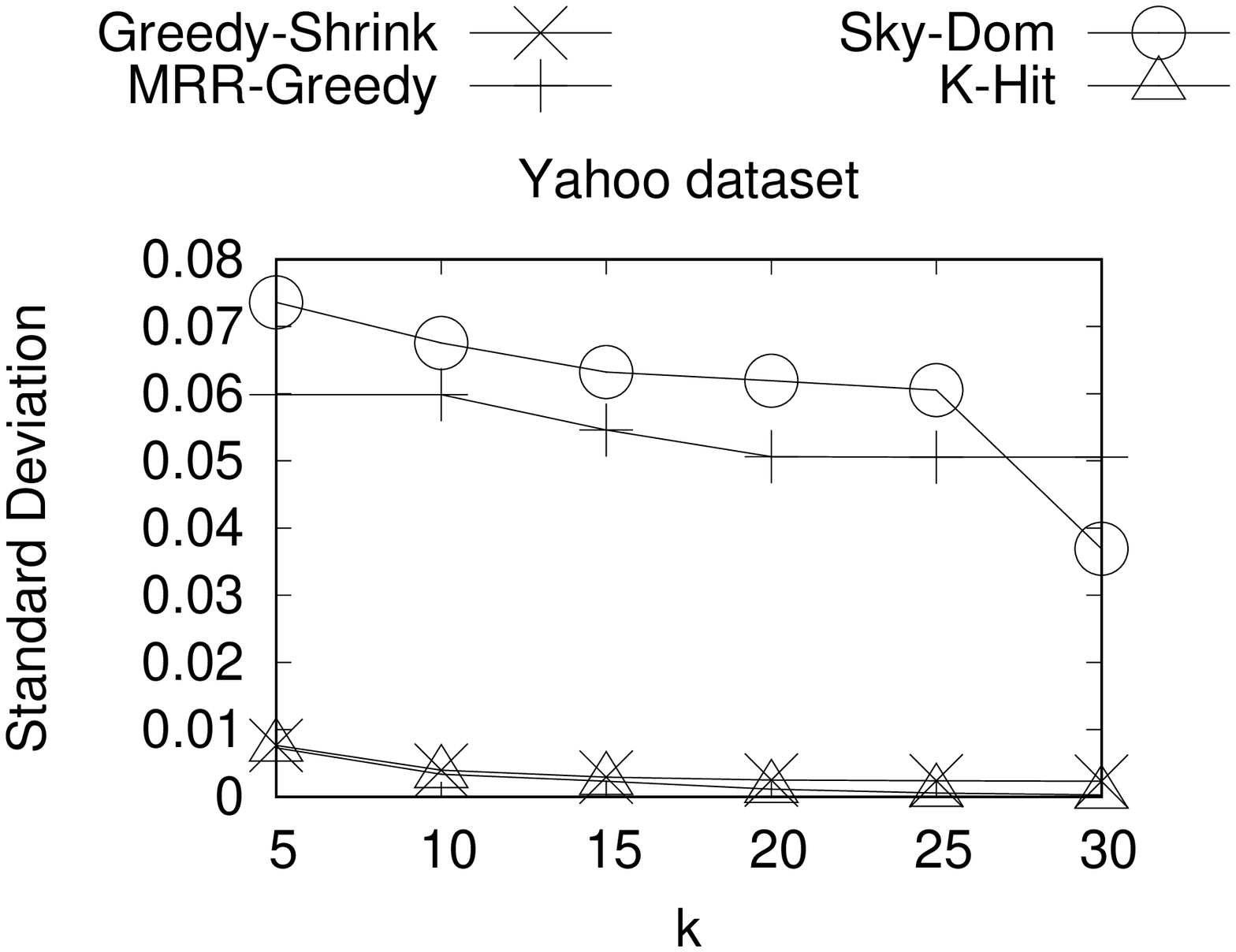}
&
  \includegraphics[width=0.4\columnwidth]{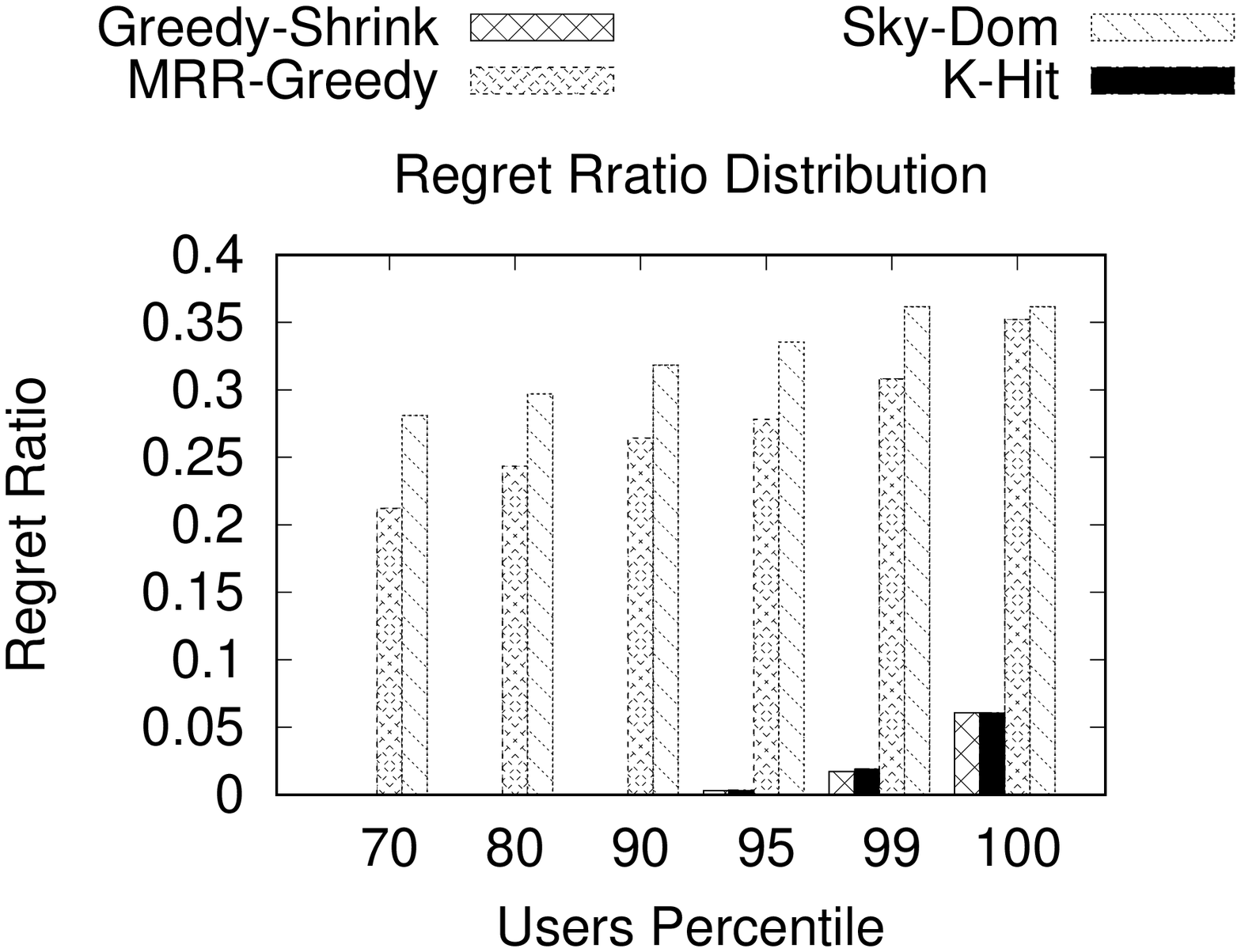}
\end{tabular}
  \caption{Effect of $k$ on standard deviation of regret ratio (left) and the distribution of the regret ratio among the users (right) on Yahoo! Dataset}
  \label{fig:yahooDataset:dist}
  \vspace{-0.4cm}
\end{figure}

\begin{figure*}[ht] 
\begin{minipage}{.65\textwidth}
  \centering
  \includegraphics[width=1\textwidth]{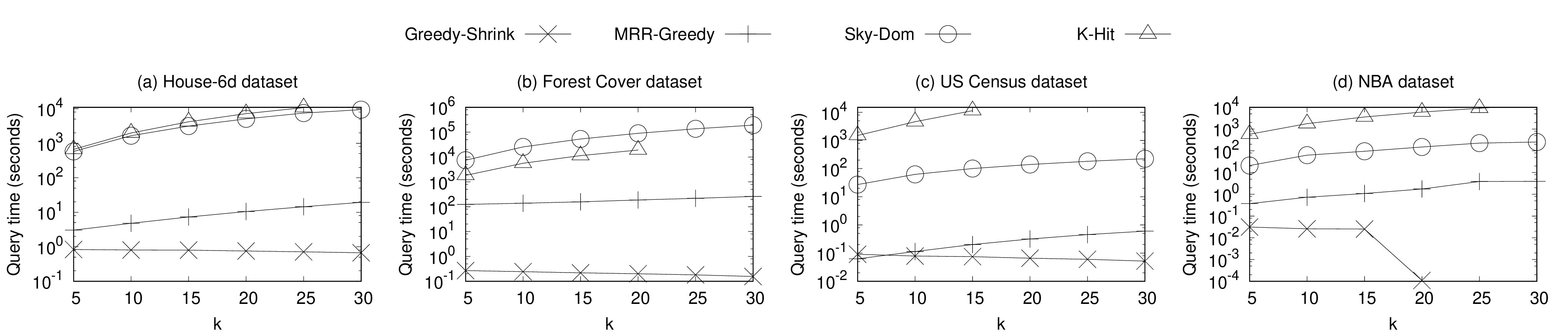}
  \caption{Effect of $k$ on query time of real datasets}
  \label{fig:realDataset:qt}
\end{minipage}%
\begin{minipage}{.35\textwidth}
  \centering
  \includegraphics[width=1\textwidth]{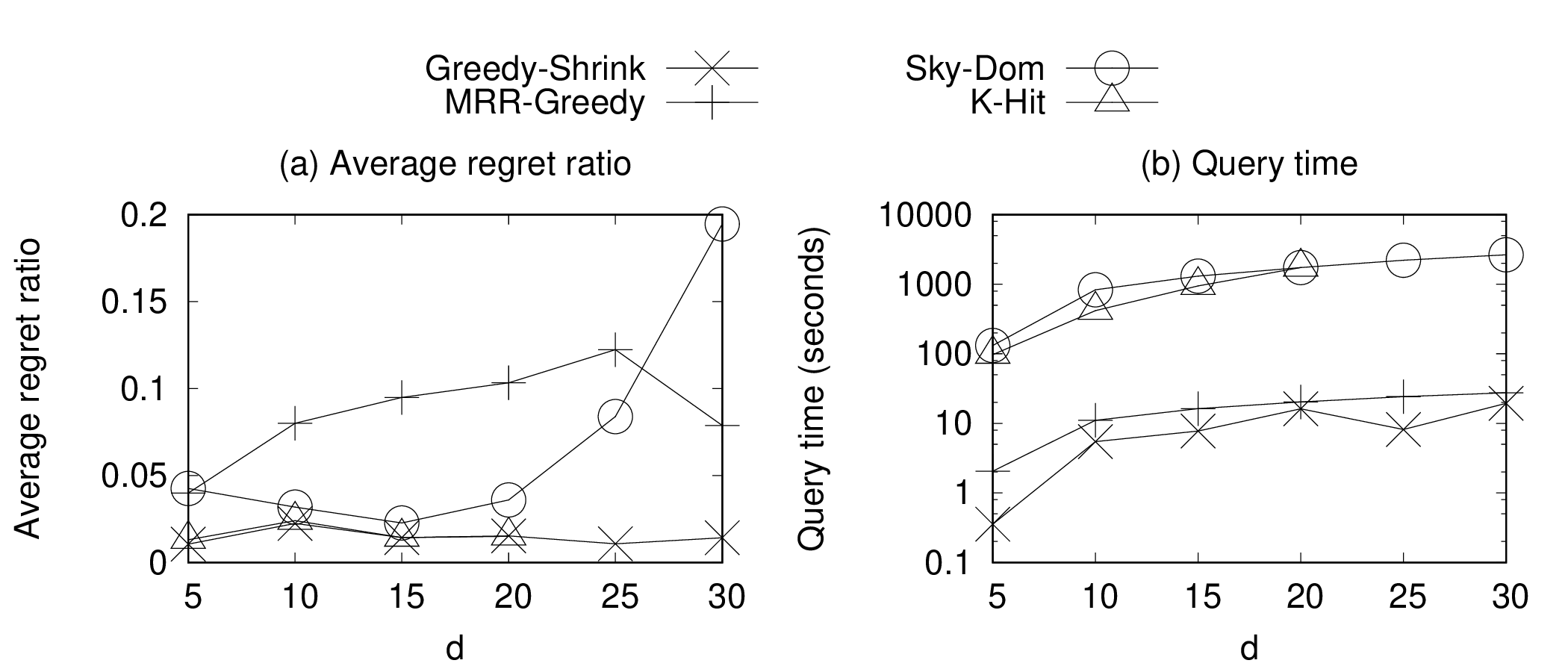}
  \caption{Effect of $d$ on Synthetic Datasets}
  \label{fig:synthDataset:d}
\end{minipage}%
  \vspace{-0.4cm}

\end{figure*}

\begin{figure*}[ht] 
\begin{minipage}{.65\textwidth}
  \centering
  \includegraphics[width=1\textwidth]{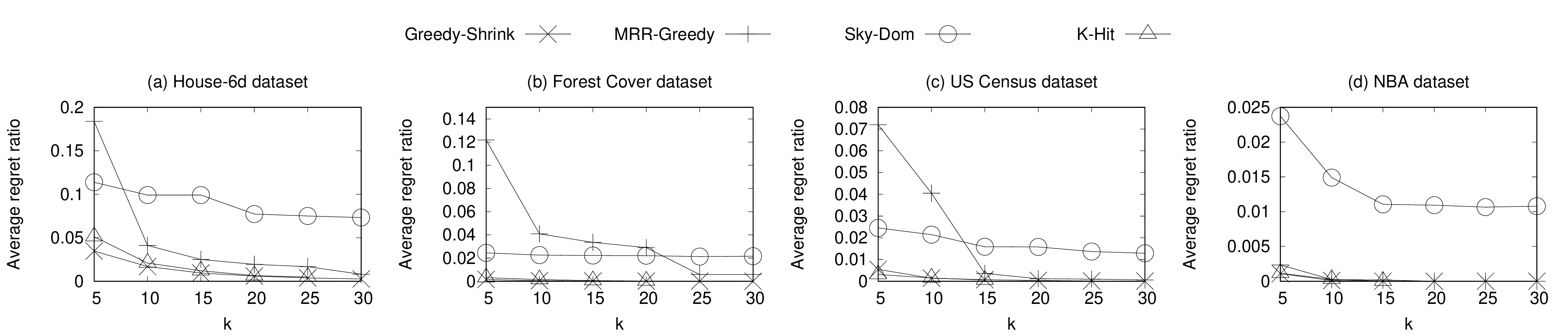}
  \caption{Effect of $k$ on average regret ratio of real datasets}
  \label{fig:realDataset:arr}
\end{minipage}%
\begin{minipage}{.35\textwidth}
  \centering
  \includegraphics[width=1\textwidth]{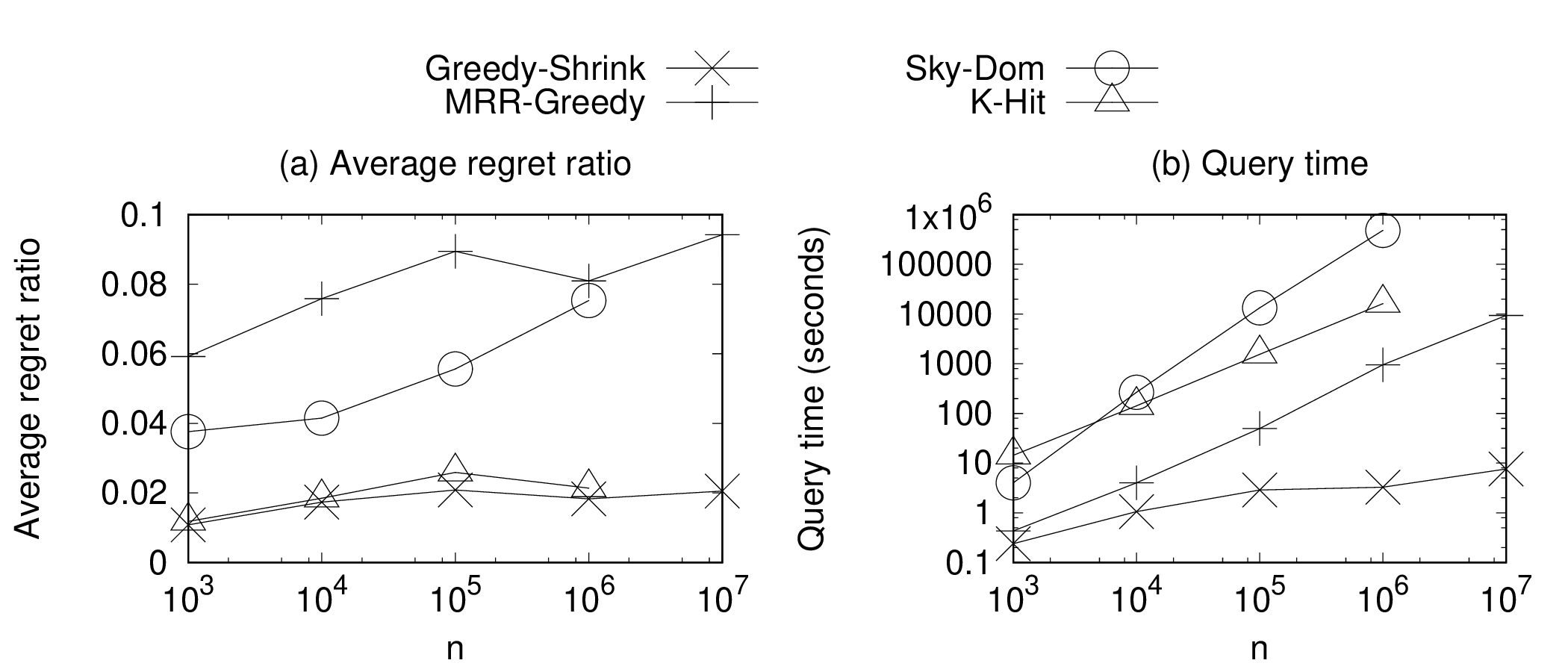}
  \caption{Effect of $n$ on Synthetic Datasets}
  \label{fig:synthDataset:n}
\end{minipage}%
  \vspace{-0.4cm}
\end{figure*}

\subsection{Experiments for Average Regret Ratio}
\label{subsec:exp:arrResult}

There are two types of datasets in our experiments, namely \emph{real datasets}
and \emph{synthetic datasets}.

There are two categories of real datasets.
The first category contains a
dataset called the Yahoo!music dataset (http://webscope.sandbox.yahoo.com/catalog.php?datatype=c) which
was used for KDD-Cup 2011 and contains the ratings provided by users for different songs.
In this dataset, following \cite{wong:kHit}, we adopted
machine learning techniques to learn a non-uniform distribution of non-linear utility functions.

The second category of real datasets contains
the four datasets commonly used in the existing
studies for skyline queries and top-k queries, namely
\emph{Household-6d} (http://www.ipums.org), \emph{Forest Cover} (http://kdd.ics.uci.edu), \emph{US Census} (http://kdd.ics.uci.edu)
and \emph{nba} (http://www.basketballreference.com).

The synthetic datasets were generated by the synthetic dataset generator \cite{skyline}.
Unless otherwise stated, in these synthetic datasets, we set $n$ to 10,000 and $d$ to 6.
Besides, in these synthetic datasets, the utility functions used are linear and
the distribution of the utility functions, $\Theta$, is uniform.

Since our competitive algorithm {\scshape{Sky-Dom}}
has a large execution time, we sampled 100,000 data points from the original
datasets of \emph{Forest Cover} and \emph{US Census} to obtain the resulting datasets
used in our experiments.
In these datasets, the utility functions used are linear and
their distribution is uniform.
The number of dimensions and the
data size of each of these real datasets can be found in
Table~\ref{tab:realDatasetStat}.

We compared our proposed algorithms called {\scshape{Greedy-Shrink}} and {\scshape{DP}} (described in Section \ref{sec:algo} and \ref{sec:twoDimAlgo})
with the 3 existing algorithms, namely {\scshape{MRR-Greedy}}~\cite{Nanongkai:regret}, {\scshape{Sky-Dom}}~\cite{Lin:exp} and {\scshape{K-Hit}}~\cite{wong:kHit}.
{\scshape{MRR-Greedy}} is the greedy algorithm~\cite{Nanongkai:regret} designed to find the solution
set based on maximum regret ratio.
{\scshape{Sky-Dom}}
is an algorithm in \cite{Lin:exp}
which selects $k$ points that together dominate the most number of points in the skyline of a dataset.
This algorithm was also included in the experimental results of
\cite{Nanongkai:regret}. {\scshape{K-Hit}} is a top-$k$ algorithm proposed by \cite{wong:kHit} that uses a probabilistic approach for selecting $k$ points.

We evaluated the algorithms with the four measurements, namely
the average and standard deviation regret ratio of the set returned by an algorithm, the distribution of regret ratio of the users and 
the query time of an algorithm. 
The query time of an algorithm corresponds to its execution time excluding the preprocessing step.
For example, the query time of {\scshape{Greedy-Shrink}} is its execution time
excluding the processing time (i.e., the time of finding the skyline of the dataset and the time of finding the best point of each of the \textit{N} sampled utility functions). The standard deviation and the distribution of the regret ratio are calculated using the same sampling method as average regret ratio. That is, the standard deviation is the standard deviation of regret ratio of the sampled utility functions. The distribution of the regret ratio of the users is calculated, similar to \cite{AKS+17}, using the regret ratio at different percentiles of users based on the sampled users.
The default value of $k$ is set to 10. In the $k$-hit
algorithm, 
we set parameters $\epsilon$ and $\delta$ to be 0.1 such that 
the setting matches the error and confidence parameter for sampling in {\scshape{Greedy-Shrink}}.

In this section, we first use a 2-dimensional dataset to compare our algorithms, {\scshape{DP}} and {\scshape{Greedy-Shrink}},
with other existing algorithms in Section~\ref{subsubsec:exp:comparison2d} and compare the quality of the solutions with the optimal solution.
Then, we present the experimental results on real datasets (Section~\ref{subsubsec:exp:experimentalResultRealDataset})
and synthetic datasets (Section~\ref{subsubsec:exp:experimentalResultSyntheticDataset}). We also performed experiments to compare the quality of our solution with an optimal solution in higher dimensions (found by a brute force approach), as well as the impact of $\epsilon$ on the solution quality, but we omit the results here as they were similar to what is presented here. The results can be found in 
\ifx\techReport\undefined
our technical report \cite{techreport}.
\else
Section \ref{appendix:exp}.
\fi

\subsubsection{Comparison with {\scshape{DP}} in a 2-dimensional dataset}
\label{subsubsec:exp:comparison2d}
In this experiment, we evaluate the performance of {\scshape{DP}} and compare the quality of the solution of the other algorithms with the optimal solution of {\scshape{DP}}.
Since {\scshape{DP}} works on a 2-dimensional dataset only,
in this experiment, the number of dimensions used is 2.
We created a synthetic dataset (following the same procedure as the other synthetic datasets in our experiments) with dimensionality equal to 2 and 10,000 points. 

Figure~\ref{fig:DP:k}(a) and (b)
shows that
overall, {\scshape{Greedy-Shrink}} and {\scshape{K-Hit}} return an average regret ratio close to the optimal value,
 but {\scshape{MRR-Greedy}} and {\scshape{Sky-Dom}} return poor approximations to the optimal solution, especially when $k$ is a large value. 
Figure~\ref{fig:DP:k}(c) shows that 
the query times of all of the algorithms are relatively small, with {\scshape{DP}} being among the highest as expected.

\subsubsection{Experimental Results on Real Datasets}
\label{subsubsec:exp:experimentalResultRealDataset}

We conducted experiments on the first-type and the second-type real datasets.

\smallskip
\noindent\textbf{First-Type Real Dataset:}
Following the experimental setup of {\cite{wong:kHit}},
we learnt the distribution $\Theta$ of the utility functions as follows. Firstly, note that we are given a dataset of ratings of different songs (data points) by different users, and that no quantitative information is available about each song except the ratings of some of the users for the songs. We can see the ratings as the utility score of a user from a point. However since not all the points are rated by all the users, we need to infer the utility score of each user for the points they have not rated. For this we use a matrix factorization technique {\cite{matrix}}, using which we can estimate the utility score of each user from each data point. Finally, to infer the probability distribution, we use a Multivariate Gaussian Mixture Model with 5 mixture models to learn the distribution of the utility functions from the utility functions obtained using the matrix factorization method. In our calculation of average regret ratio, we sample users from the Gaussian Mixture Model. In this dataset, we have 8,933 data points in the database.

Figure~\ref{fig:yahooDataset:k} shows how the change in $k$ affects the average regret ratio and the query time of each algorithm. As it can be observed, {\scshape{Greedy-Shrink}} and {\scshape{K-Hit}} work well on this real dataset, returning a very small average regret ratio. 
{\scshape{MRR-Greedy}}'s average regret ratio is relatively high. Moreover, both {\scshape{Greedy-Shrink}} and {\scshape{MRR-Greedy}} are very fast in practice,
 but {\scshape{Sky-Dom}} and {\scshape{K-Hit}} have a larger query time.
 
 More interestingly, Figure~{\ref{fig:yahooDataset:dist}} shows the standard deviation and distribution of regret ratio among the users. Both {\scshape{MRR-Greedy}} and {\scshape{Sky-Dom}} have a larger standard deviation compared with {\scshape{Greedy-Shrink}} and {\scshape{K-Hit}}, and their regret ratio of the users is larger at all the user percentiles. This can be attributed to the fact that both {\scshape{MRR-Greedy}} and {\scshape{Sky-Dom}} do not take into account the distribution of the utility functions which can result in more probable users having larger regret ratio values compared with  {\scshape{Greedy-Shrink}} and {\scshape{K-Hit}} that take into account the distribution. 

\smallskip
\noindent\textbf{Second-Type Real Datasets:}
For the datasets used here, unlike the First-Type Real Datasets, we do not have any information regarding the distribution of the utility function. Thus we assumed that the utility functions are distributed uniformly. Figure~\ref{fig:realDataset:arr} shows the average regret ratios of the solutions returned
by different algorithms based on datasets, Household-6d, Forest Cover, US Census and NBA.
{\scshape{Greedy-Shrink}} has the smallest average regret ratio among all the algorithms,
and {\scshape{K-Hit}} has a slightly larger average regret ratio.
 However, {\scshape{Sky-Dom}} algorithm does not work well on real datasets and
 returns an average regret ratio much larger than the other algorithms. 
Furthermore, the average regret ratio of the points returned by {\scshape{Sky-Dom}} does not change significantly when the number of points returned increases.

Figure~\ref{fig:realDataset:qt} shows the query times of the solutions returned
by different algorithms based on datasets, Houshold-6d, Forest Cover, US Census and NBA.
{\scshape{Greedy-Shrink}} has the smallest query times.
 On the other hand, {\scshape{Sky-Dom}} and {\scshape{K-Hit}} took a very long time to return solutions. The better query performance of {\scshape{Greedy-Shrink}} compared with {\scshape{MRR-Greedy}} can be attributed to the practical improvements we made to the algorithm, as discussed in Section {\ref{subsec:alg:detailedStep}}. Using the improvements, for each calculation of average regret ratio, we only need to recompute the regret ratio of about 1\% of the users on average per iteration and we only need to consider 68\% of the points per iteration.

We also performed experiments on the standard deviation and distribution of regret ratio among the users, and the results were similar to the results for the first type real datasets. We experimented with increasing the sample size for calculating the distribution of regret ratio among the users to $N = 1,000,000$, but there was no significant change in the distribution of the regret ratio among users. We omit these results here for brevity, and they could be found in 
\ifx\techReport\undefined
our technical report \cite{techreport}.
\else
Section \ref{appendix:exp}.
\fi

\subsubsection{Experimental Results on Synthetic Datasets}
\label{subsubsec:exp:experimentalResultSyntheticDataset}

We conducted experiments on synethetic datasets for scalability test.
We varied $d$ and $n$ to see the scalability of the proposed algorithm. The results are shown in Figures ~\ref{fig:synthDataset:d} and ~\ref{fig:synthDataset:n}. The figures shows that our algorithm is scalable and is capalbe of handling large values of $n$ and $d$.

\noindent\textbf{Summary.} Overall, the average regret ratios of the {\scshape{Greedy-Shrink}} and {\scshape{K-Hit}} algorithms are always smaller than both of the other algorithms and are less critically affected by the change in dimensionality and the size of the database, while both have a lower standard deviation and provide a lower regret ratio for the majority of the users. Moreover, {\scshape{Greedy-Shrink}} has the lowest query time but {\scshape{Sky-Dom}} and {\scshape{K-Hit}} become impractical for large datasets.

\section{Related work}\label{sec:rel}
There are a lot of existing studies in the literature about finding
the ``best'' point of a user when the utility function of this user is given.
One representative branch is top-$k$ query processing~\cite{SIC07,topk:Ilyas,HL12}.
Given a utility function of a user, a top-$k$ query is to return the $k$ points with the greatest utilities with respect
to the utility function. However, it requires the user to provide the exact utility function.

On the other hand, there are also many studies about finding a set of candidates for the ``best'' point of a user
when the utility function of this user is unknown.
There are two categories.
The first category is that there is no information about the utility function of any user.
The first type under the first category is \emph{skyline queries} \cite{skyline}
which are to find a set of points which are not dominated by any other points in the database.
As pointed out by many existing studies, the answer of skyline queries could possibly contain a lot of points and thus the output size is uncontrollable, which is not user-friendly
to a user.
The second type is some variants of \emph{skyline queries} \cite{Lin:exp,topk:Papadopoulos,topk:Goncalves}
which aims at overcoming the drawback of existing skyline queries by restricting the output size to be at most a user parameter $k$.
Some examples are a representative skyline query \cite{Lin:exp}, a dominating skyline query \cite{topk:Papadopoulos}
and a top-$k$ skyline query \cite{topk:Goncalves}.

The third type is \emph{$k$-regret queries} \cite{ANZ+17,AKS+17,CLW+17,KBL15,NLS+12,chester:regret,regret:Peng,Nanongkai:regret}, recently proposed queries in the database community,
which could address both the issue of top-$k$ query processing (i.e., requiring an exact utility function of
a given user) and the issue of skyline queries (i.e., returning an output set with an uncontrollable size).
That is, a $k$-regret query does not require a user to give an exact utility function of a given user
and returns the output set with a controllable size.

Specifically, it
is to return a set $S$ of $k$ points such that the \emph{maximum regret ratio} for set $S$ is minimized. 
Here, the maximum regret ratio for set $S$ is defined to be $\max_{s \in F} rr(S, f)$.
It was shown in \cite{AKS+17,CLW+17,chester:regret} that solving a $k$-regret query is NP-hard.
Existing studies about $k$-regret queries focused on improving the efficiency of a proposed algorithm
and improving the quality of the result (i.e., reducing the maximum regret ratio for the answer set).
As we described in Section~\ref{sec:intro}, optimizing the \emph{worst-case} scenario (which corresponds to $k$-regret queries)
is not useful in some applications. Instead, optimizing \emph{average-case} scenario (which corresponds to our FAM problem)
is more useful since it consider the expectations of different users.

The second category is that there is some information about the utility functions for the whole population including different users (not a particular user) which corresponds to the distribution $\Theta$ of the utility functions for the whole population. The first type under the second category is $k$-hit queries \cite{wong:kHit}. Recently, \cite{wong:kHit} considered the distribution $\Theta$ of the utility functions for the whole population and assumed that the probability that a user has a utility function $f$ in $F$ follows distribution $\Theta$. 

Specifically, \cite{wong:kHit} proposed a \emph{$k$-hit query} which is to find a set $S$ of $k$ points such that the \emph{probability} that at least one point
in $S$ is the best point of a user is maximized.
The answer to this query becomes less convincing if 
we care about not only users which regard the points in the answer set as the best points
but also users which do not regard the points in the answer set as the best points.
This is because each point in the answer set is exactly the best point of a certain number
of users based on $\Theta$. Thus, it does not consider any users which do not regard
the points in the answer set as the best points.
However, our FAM problem considers the preferences from
both the users who regard the points in the answer set of the FAM problem as the best points
and the users who do not.
Roughly speaking, each point in the answer set is a point which is ``close'' to the
best point of any user in the population. Thus, in practice, the result of our FAM problem
is more convincing than the result of the $k$-hit query.

The second type under the second category is the FAM problem~\cite{zeighami}
which was published in SIGMOD 2016 Undergraduate Research Competition.
\cite{zeighami}
 first studied the FAM problem
and proposed a greedy algorithm for this problem.
However, there are the following differences between \cite{zeighami} and this paper.
Firstly, \cite{zeighami} formulated the FAM problem without any experimental
justification about why the FAM problem is better than existing queries, while we study this problem with experimental justification.
Secondly, \cite{zeighami} presented some results without any proof but
this paper includes the results together with detailed proofs.
Thirdly, \cite{zeighami} did not include the following results which could be found in this paper only:
firstly, the NP-hardness result, 
secondly,  the dynamic programming
algorithm for the FAM problem when the dataset contains two dimensions,
and thirdly,
the comprehensive experimental results including the justification about why the FAM problem
is better than existing queries and including the comparison with many existing related algorithms.

\section{Conclusion}\label{sec:con}
In this paper, we considered the problem of selecting a number of representative points from a database. The problem is concerned with the happiness, or utility, of the users who see the selected points instead of the whole database. Since we do not know each user's utility function, we try to select points that maximize the expected happiness of the users. Therefore, we aim at minimizing the average regret ratio of a user when he or she sees the set of $k$ selected points, instead of the whole database. We discussed a sampling approach for the calculation of the average regret ratio and gave a greedy approximation algorithm to find a solution set in polynomial time, based on the supermodularity of the average regret ratio. We also provided a dynamic programming algorithm to solve the problem optimally in the 2-dimensional case. Methods to improve the performance of the algorithm empirically were also discussed and extensive empirical studies were performed.

\bibliographystyle{abbrv}
\bibliography{arr}

\ifx\techReport\undefined
\else
\appendix
\subsection{Handling the Case of Countable $F$}\label{appendix:countable}
Here, we discuss how we can handle the case when the distribution of the utility functions is discrete or the set $F$ is countable. 

In the case when $F$ is countable, we are no longer dealing with a continuous space of utility functions. Hence, our definition of average regret ratio needs to be altered. In our previous definition (provided in Section ~\ref{sec:def}), average regret ratio was considered as a continuous random variable, but when $F$ is countable, regret ratio will be a discrete random variable whose domain is $F$. As a result, in such a setting, average regret ratio will be the expected value of this discrete random variable and can be defined as follows.

\begin{definition}[Average Regret Ratio (Discrete Space)]
	Let $F$ be a countably finite set of of users with the probability density function $\eta(.)$
	and $S$ be a subset of $D$.
	The \emph{average regret ratio} of $F$
	for $S$, denoted by $arr(S)$, is defined to be $\sum_{f\in F} rr(S, f)\eta (f)$.
	
\end{definition}

When $F$ is countable, we call the distribution of utility functions discrete. If the set $F$ is countably finite, we can calculate the exact value of average regret ratio, while sampling methods may still be useful if the set $F$ is large. Next, we discuss the case of a countably finite $F$ in more details. 

\subsubsection{Discrete Distribution of Finite Utility Functions}
The calculation of the average regret ratio in this case is simple. We need to calculate $arr(S) = \sum_{f\in F} rr(S, f)\eta (f)$. So, for each user, we need to calculate the regret ratio and then sum all the regret ratios weighted by their probabilities. For example, we can calculate the average regret ratio for the users shown in Table \ref{tab:example:hotelUtil} (assuming uniform distribution on the utility functions, i.e. 0.25 probability for each user) for the set $S = $\{Intercontinental, Hilton\} as $arr(S) = rr(S, Alex)\times 0.25 + rr(S, Jerry)\times 0.25 + rr(S, Tom) \times 0.25 + rr(S, Sam) \times 0.25$.

\textit{An Example on Sampling.} We can also calculate the average regret ratio using sampling. To do so, we sample a number of utility functions from the 4 different utility functions in Table \ref{tab:example:hotelUtil} (assuming uniform distribution). Consider the sample size, $N$, equal to 10. We randomly select 10 utility functions from the the set of utility functions. Imagine that the outcome is the set $F_N = $ \{Alex, Alex, Sam, Tom, Alex, Tom, Jerry, Jerry, Sam, Sam\}. Then, we can calculate the average regret ratio for these 10 utility functions from the solution set $ S = $\{Intercontinental, Hilton\}. We get $arr(S) = \big( rr(S, Alex)\times 3 + rr(S, Jerry)\times 2 + rr(S, Tom) \times 2 + rr(S, Sam) \times 3\big)/10$.

\subsection{Supplementary Experimental Results}\label{appendix:exp}
\begin{table}
\centering\begin{tabular}{|c | c | c |} \hline
  $\epsilon$  &$\sigma$ & $N$    \\ \hline
  0.01  &	0.1 & 69,077 \\\hline
  0.001  &	0.1 & 6,907,755 \\\hline
  0.0001  &	0.1 &  690,775,528 \\\hline
  0.01  &	0.05 & 89,871 \\\hline
  0.001  &	0.05 & 8,987,197 \\\hline
  0.0001  &	0.05 &  898,719,682 \\\hline
\end{tabular}
\caption{Sample size $N$ for some chosen values of $\epsilon$ and $\sigma$}
\label{table:5}
\label{tab:samplingSize}
\vspace*{-0.5cm}
\end{table}
\begin{figure}
    \centering
	\includegraphics[width=1\columnwidth]{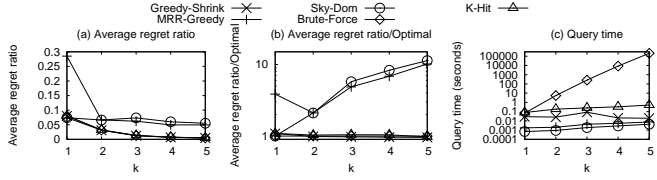}
	\caption{Effect of $k$ on small real sampled dataset}
	\label{fig:sampledDataset:k}
\end{figure}
\begin{figure}
	\includegraphics[width=0.45\textwidth]{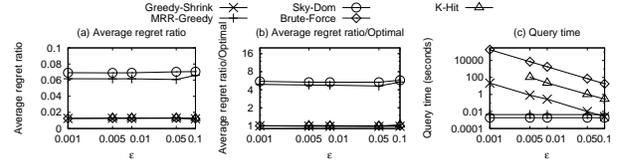}
	\caption{Effect of $\epsilon$ on small real sampled dataset}
	\label{fig:sampledDataset:e}
\end{figure}%
\begin{figure*}
\centering
  \centering
  \includegraphics[width=0.65\textwidth]{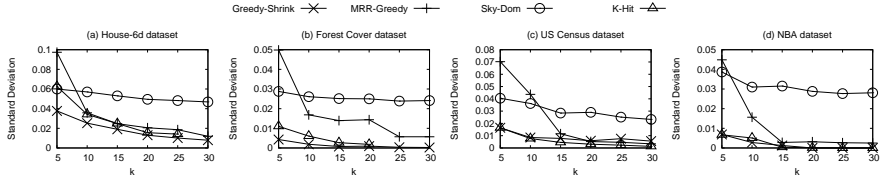}
  \caption{Effect of $k$ on standard deviation of regret ratio of real datasets}
  \label{fig:realDataset:sd}
\end{figure*}
\begin{figure*}
  \centering
  \includegraphics[width=0.65\textwidth]{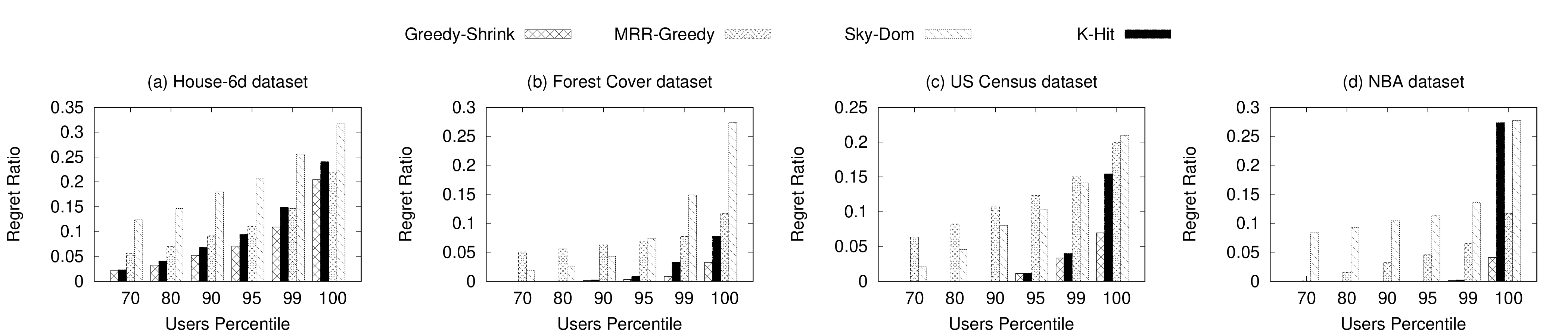}
  \caption{Regret ratio distribution of real datasets (N=10,000)}
  \label{fig:realDataset:dist}
\end{figure*}%
\begin{figure*}
  \centering
  \includegraphics[width=0.65\textwidth]{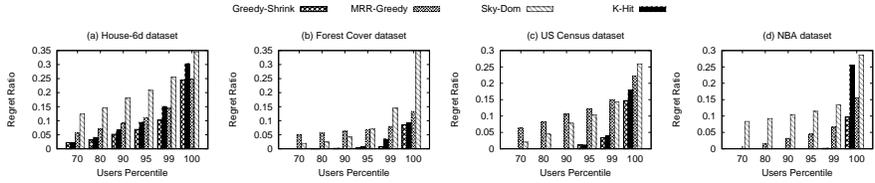}
  \caption{Regret ratio distribution of real datasets (N=1,000,000)}
  \label{fig:realDataset:distLarge}
\end{figure*}%

\subsubsection{Comparison with Brute-force Method}
\label{subsubsec:exp:comparisonWithBruteforceMethod}
In this experiment, since we need to compare with a brute-force method which is computationally expensive,
we sampled 100 points from the real dataset ``Household-6d'' to
obtain a smaller dataset. Here, we varied $k$ and $\epsilon$.

\textit{Effect of $k$.} Figure~\ref{fig:sampledDataset:k}(a) and (b) show that, {\scshape{Greedy-Shrink}} and {\scshape{K-Hit}} return an average regret ratio close to the optimal value, but the other algorithms return poor approximations to the optimal solution, especially when $k$ is a large value. 
Figure~\ref{fig:sampledDataset:k}(c) shows that 
the query times of all of the algorithms except {\scshape{Brute-Force}} are very small. 
\textit{Effect of $\epsilon$.} Our result shows that changing $\epsilon$ from 0.1 to 0.001 has a marginal effect on the average regret ratio and the quality of the solution for all the algorithms involved. Please refer to Section \ref{appendix:exp} for more details.

\textit{Effect of $\epsilon$.} Note that a smaller $\epsilon$ means more sampled utility functions used in our sampling (Table~\ref{tab:samplingSize} shows the sampling size $N$ for some chosen values of $\epsilon$ and $\sigma$).
Figure~\ref{fig:sampledDataset:e}(a) and (b) show that
$\epsilon$ (i.e., the error parameter used in our sampling) does not affect the results
returned by algorithms a lot. This is because the values used in this error parameter $\epsilon$
are also small (e.g., 0.001 and 0.1).
Figure~\ref{fig:sampledDataset:e}(c) shows that
the query times of the the {\scshape{Greedy-Shrink}}, {\scshape{Brute-Force}} and {\scshape{K-Hit}} increase when $\epsilon$ decreases, with {\scshape{Greedy-Shrink}} having the
smallest value among the three. This is because they also involve a process of calculating the average regret ratio relying
on the number of sampled utility functions. It is easy to verify that
algorithms {\scshape{MRR-Greedy}} and {\scshape{Sky-Dom}} remain unchanged with the change of $\epsilon$
(because these two algorithms do not involve any process of calculating the average regret ratio whose time cost relies
on the number of sampled utility functions).

\subsubsection{Standard Deviation and Distribution of Regret Ratio on Real Datasets}
Figure~\ref{fig:realDataset:sd} shows that in practice, both {\scshape{Greedy-Shrink}} and {\scshape{K-Hit}} have low standard deviation while {\scshape{MRR-Greedy}} and {\scshape{Sky-Dom}} have higher values, although the standard deviation decreases as more points are selected. Moreover, Figure~\ref{fig:realDataset:dist} shows that even upto 99\% of the sampled users enjoy a very low regret ratio for the sets selected by {\scshape{Greedy-Shrink}} and {\scshape{K-Hit}}, while the regret ratio of most of the users is larger for sets returned by {\scshape{MRR-Greedy}} and {\scshape{Sky-Dom}}. Note that in Figure~{\ref{fig:realDataset:distLarge}}, we increased the sample size to 1,000,000 users to reduce the possible error rate in finding the percentiles, but as the figure shows, the results are very similar to the case when the sample size is 10,000.

The results above show that even if we do not have any information regarding the distribution of the utility functions and we assume it is uniform, minimizing average regret ratio results in a better distribution of the regret ratio of the user, and in practice, the majority of the users may have lower regret ratio if we aim at minimizing average regret ratio instead of maximum regret ratio.

\subsection{Details on Improving Efficiency of GREEDY-SHRINK}\label{appendix:improve}

\textbf{Improvement~1} (Best Point Calculation). Improvement~1 is to compute the best point of a utility function efficiently with a pre-computation step. With this improvement, the efficiency of algorithm {\scshape{Greedy-Shrink}} (i.e., Algorithm~\ref{algo}) could also be improved.

Specifically, in each iteration of
algorithm {\scshape{Greedy-Shrink}} (i.e., Algorithm~\ref{algo}),
we have to compute $\arg\min_{p \in S} arr(S-\{p\})$.
Here, we need to compute the average regret ratio, $arr(\cdot)$, by using Equation~(\ref{eqn:samplingFormula}).
In the summation of this equation, we need to compute two terms, namely
$\max_{p\in D}f(p)$ and $\max_{p\in S}f(p)$, for a particular function $f$ in $F_N$.
These two terms correspond to the utilities of the \emph{best points} of a utility function in $D$ and $S$, respectively.

In particular,
we could compute the first term (i.e., $\max_{p\in D}f(p)$) in $O(n)$ time, and
the second term (i.e., $\max_{p\in S}f(p)$) in $O(|S|)$ time.
The major idea of Improvement~1 is to store the two corresponding best points
so that whenever we compute $arr(\cdot)$, we do not need to re-compute
these two terms from scratch and instead, we directly compute them based
on these two best points stored. Besides, when the solution set $S$ being
maintained by algorithm {\scshape{Greedy-Shrink}} changes, we have to update the
best point of a utility function in $S$ accordingly. In our experiments, on average, the best point of only about 1\% of the users changes per iteration of the algorithm. As a result, computation of average regret ratio in Algorithm {\ref{algo}} can be done efficiently in practice.

\textbf{Improvement 2 (Computation based on Previous Iteration).}
Improvement~2 is to
 re-use the computation obtained
at one of the previous iterations for the current iteration.
Since a lot of computation could be re-used in algorithm {\scshape{Greedy-Shrink}} with Improvement~2,
 the efficiency of algorithm {\scshape{Greedy-Shrink}} could also be improved.

Consider an iteration of the algorithm.
Consider a solution set $S$ maintained by the algorithm just before the beginning of this iteration.
Just before the beginning of each iteration, for each point $p$ in $S$, the algorithm computes
$arr(S - \{p\})$. Then, it finds the point $p_o$ with the smallest value
of $arr(S - \{p_o\})$.

Let $v_{p, S}$ be $arr(S - \{p\})$ for each $p \in S$.
We call $v_{p, S}$ to be the \emph{evaluation value} of point $p$
based on set $S$.

Let $S_{curr}$ be the solution set maintained by the algorithm
just before the beginning of the current iteration.
Let $S_{prev}$ be the solution set maintained by the algorithm
just before the beginning of the previous iteration.
Note that $S_{prev}$ is exactly equal to $S_{curr} \cup \{p_o\}$
where $p_o$ is the point with the smallest evaluation value based on $S_{prev}$
(among all points in $S_{prev}$).

We have the following lemma.
\begin{lemma}
For each point $p \in S_{curr}$,
$$
     v_{p, S_{curr}} \ge v_{p, S_{prev}}
$$
\label{lemma:lowerBound}
\end{lemma}

Lemma~\ref{lemma:lowerBound} suggests that
the evaluation value based on the previous iteration (i.e., $v_{p, S_{prev}}$)
could be regarded as the lower bound of the evaluation value
based on the current iteration.
Next, we describe how we use this ``lower bound'' property to speed up our computation
with the following lemma.

\begin{lemma}

Let $p_o$ be the point with the smallest evaluation value based on $S_{curr}$ (among all points in $S_{curr}$).

Consider a point $p$ in $S_{curr}$.
Let $v$ be the evaluation value of $p$ based on $S_{curr}$.

(1) Each point  in $S_{curr}$ whose evaluation value based on $S_{prev}$
is larger than $v$ is not equal to $p_o$.
(2)
If there is no point in $S_{curr}$ whose evaluation value based on $S_{prev}$
is smaller than $v$, then $p_o$ is equal to $p$.
\label{lemma:lowerBoundUsage}
\end{lemma}

Lemma~\ref{lemma:lowerBoundUsage} suggests that whenever we have
a point $p$ in $S_{curr}$ and compute its evaluation value $v$ based on $S_{curr}$, we just need to process
each point in $S_{curr}$ whose evaluation value based on $S_{prev}$ 
is smaller than or equal to $v$
(because each point in $S_{curr}$ whose
evaluation
value based on $S_{prev}$
is larger than $v$ is not equal to	
the point we want to find, i.e., the point with the smallest
evaluation value based on $S_{curr}$).
Besides, if there is no point in $S_{curr}$ whose evaluation value based on $S_{prev}$
is smaller than $v$, then $p$ is the point we want to find.

Consider an iteration.
Let $S_{curr}$ be the solution set maintained just before the beginning of the iteration.
In this iteration, we have to find the point $p$ with the smallest
evaluation value based on $S_{curr}$.
A straightforward implementation is to compute the evaluation values
of \emph{all} points (in $S_{curr}$) based on $S_{curr}$ and find the point $p$ with the smallest
evaluation value based on $S_{curr}$.

With this ``lower bound'' property, we do not need to compute
the evaluation values of  \emph{all} points $p$ (in $S_{curr}$) based on $S_{curr}$.
Specifically, we do the following.
\begin{itemize}
   \item Firstly, just before the beginning of the first iteration,
   we compute the evaluation values of all points in $S_{curr}$ based on $S_{curr}$ 
   and sort all points in ascending order of this evaluation value.
   Thus, we maintain a sorted list $L$ where each point is associated with the evaluation value just computed.
   Then, we find the point $p_o$ with the smallest evaluation value. 
   We set $S_{prev}$ to be $S_{curr}$ and
   then remove $p_o$ from $S_{curr}$.
   Note that all computed evaluation values in this iteration
   become the old/previous evaluation values in the next iteration.
   \item Secondly, for each non-first iteration,
   we do the following.
   \begin{itemize}
      \item For each point $p$ in $S_{curr}$, we introduce a flag variable $\theta_p$
      to indicate whether the value stored in the list $L$ (computed based on one of previous iterations)
      has been updated to the value based on the current iteration.
      Initially, $\theta_p$ is set to ``false''.
      \item We find the point $p$ in $S_{curr}$ with the smallest ``associated'' evaluation value in the list $L$.
      \begin{itemize}
        \item If $\theta_p$ is ``false'', we do the following.
      Then, we compute the evaluation value of $p$ based on $S_{curr}$ and set $\theta_p$ to ``true''.
      After that, we remove $p$ from the top of the list $L$ and
      re-insert it to $L$ based on its newly computed value such that the ascending order of
      the values in the list $L$ is still maintained. $p$ is now associated with this newly computed evaluation value.
       \item If $\theta_p$ is ``true'', we know that $p$ is the point with the smallest evaluation value based on $S_{curr}$ (among all points in $S_{curr}$) (based on Lemma~\ref{lemma:lowerBoundUsage}).
      \end{itemize}
   \end{itemize}
\end{itemize}

In our experiments,
for each iteration, on average, about 32 percent of the points in the list $L$
do not need to be re-computed with the value based on the current iteration.
Thus, a lot of computations could be saved.

\subsection{Proofs of Lemmas/Theorems}
\smallskip\noindent
\textit{Proof of Theorem~\ref{thm:NPHardness}.}
To prove that FAM is NP-hard we show that the Set Cover decision problem, referred to as SC, can be reduced to FAM in polynomial time. The SC problem is to determine from a universe $U$ and a collection of its subsets $S$, whether there exists at most $k$ sets in $S$ whose union is equal to the set $U$. SC is proven to be NP-Complete by \cite{Karp1972} and is formally defined as follows.

\indent \textit{Set Cover Problem Definition.} Given a set of items $U$, a collection of its subsets $T$, where $T \subseteq 2^U$, and an integer $k$, determine whether there exists a set $S = \{S_1, S_2, S_3, ..., S_k\}$, $S \subseteq T$ and $\left\vert S \right\vert = k$ where $\cup_{1\leq i \leq k} S_i = U$.

An instance, ${I_{SC}}$, of SC is defined by the sets $U$ and $T$ and an integer $k_{SC}$ and an instance, $I_{FAM}$, of FAM is defined by a set $D$, a distribution of utility functions $\Theta$ and an integer $k_{FAM}$. Our reduction, $\mathcal{R}$, takes the instance ${I_{SC}}$ of SC and outputs the instance $I_{FAM}$ of FAM in polynomial time as follow. Without loss of generality, we focus on the non-trivial instances, that is, the instances for which all elements in $U$ are present in at least one set in $T$. 

For ease of notation, we impose an arbitrary ordering on the elements of $U$ and $T$ so that the $i$-th element of $U$, denoted by $u_i$, refers to the $i$-th element present in that ordering and the $i$-th element of $T$ denoted by $t_i$, refers to the $i$-th set in $T$. Furthermore, let $N = |U|$ and $n = |T|$.

In the reduction, we create $D$ as a dataset containing $n$ datapoints, such that each datapoint corresponds to one set in $T$.  Furthermore, to define $\Theta$, we first define $N$ continuous subspaces of utility functions, called $F_i$ for $1 \leq i \leq N$. We let $\Theta$ be any distribution on the union of these spaces such that any range of utility functions that is a subspace of any $F_i$ has a non-zero probability and any range of utility functions that is a subspace of $F - \cup_i F_i$ has a zero probability, where $F$ is the space of all utility functions. 

To define each $F_i$, first note that because $D$ has $n$ records, $F$ is an $n$-dimensional space. Secondly, let $U_i$ be defined as the set of indexes (in the imposed arbitrary ordering) of the sets $t$ in $T$ such that $u_i \in t$, i.e. $U_i$ denotes all the sets that the element $u_i$ belongs to. Then, $F_i$ is defined as $\{(v_1, v_2, ..., v_n)| v_j = c\: \forall j\in U_i,\  v_j=0\: \forall j \notin U_i,\  c > 0\}$. For example, if an element $u_i$ exists in only the first two sets in $T$, then $F_i$ will be the set of utility functions with $n$-dimensional utility vectors of the form $f = (c, c, 0, 0, 0, ..., 0)$ for positive values of $c$.

In addition, we let $k_{FAM} = k_{SC}$. The solution to $\mathcal{I_{SC}}$ is \textit{yes} if and only if The solution, $S$, of the created instance ${I_{FAM}}$ has average regret ratio equal to $0$. 

\begin{lemma}\label{redPol}
	There exists an implementation of the reduction, $\mathcal{R}$, described above that runs in polynomial time in the input size, i.e polynomial in $\left\vert U\right\vert$ and $\left\vert T\right\vert$.
\end{lemma}

\textit{Proof.} $\mathcal{R}$ involves creating the database $D$ and the distribution $\Theta$. Creating $D$ requires one linear scan of all the sets in $T$ to create a point in $D$ for each of them. Creating each $\Theta$ requires finding what utility functions belong to each $F_i$ which requires finding the sets $U_i$. This can be done with a linear scan of all the sets $t$ in $T$ to see whether an element $u_i$ belongs to each set or not which can be done in $O(|T|\times|U|)$. Then we can represent $\Theta$ with any probability distribution function that has non-zero values on the domain that overlaps with each $F_i$ and zero values anywhere else. Thus, the time complexity of the reduction will be $O(|T| + |T|\times|U|)$ which is polynomial in $|T|$ and $|U|$. 

\hfill\ensuremath{\square}

Next, we show that the reduction is correct. For this we first provide the following lemma that will be used later.

\begin{lemma}\label{redCorr2}
When an instance of SC, ${I_{SC}} = (U, T, k_{SC})$, is reduced to an instance of  FAM, ${I_{FAM}} = (D, \Theta, k_{FAM})$, then, a set $T^*$, $T^* \subseteq T$ is a set cover of $U$ if and only if it corresponds to a set with average regret ratio equal to zero in ${I_{FAM}}$.
\end{lemma}

\textit{Proof.} First recall that there is a one to one correspondence between the points in $D$ and the elements of $T$ under the reduction. Let the set of points in $D$ that correspond to the items of $T^*$ be denoted by $S$.

\textbf{Only if. }If $T^*$ is a set cover of $U$, then all the elements of $U$ are present in some set in $T^*$. Based on the reduction, since $T^*$ covers all the elements in $U$, then for any $i$ the utility function $f_i \in F_i$ will have utility $c$ from some point in $S$. This is because as $T^*$ covers all the elements in $U$, for any $i$, $T^*$ must contain at least a set that is in $U_i$ as $U_i$ includes the indexes of all the sets that contain $u_i$.

\textbf{If.} Let $S$ be a set with average regret ratio 0. This implies that the regret ratio of all the user in all $F_i$ sets must be zero (no other utility function contribute to the average regret ratio because of the probability distribution). Based on the definition of regret ratio, for any utility function $f$, we will have $\max_{p\in D}f(p) = \max_{p\in S}f(p)$. Furthermore, for any utility function $f$, and for any point $p \in D$, based on the reduction, $\frac{\max_{p\in S}f(p)}{\max_{p\in D}f(p)}$ can take only two possible values, either 0 or 1, since $\max_{p\in S}f(p)$ is either zero or equal to $\max_{p\in D}f(p)$. First consider two cases for any $f \in F_i$ for any $i$, that $\max_{p\in D}f(p)$ is either zero or non-zero. 

$\max_{p\in D}f(p) = 0$ means that the utility function $f$ gains utility 0 from all the points in the database. Based on our reduction, this would imply that for some $i$, the set $U_i$ must be empty or otherwise $f(p)$ would have been non-zero for the point that corresponds to the members of $U_i$. $U_i$ being empty in turn implies that no set in $T$ includes the element $u_i$. This, however, is contradictory with our assumption that all elements of $U$ are present in some set $T$. Therefore, $\max_{p\in D}f(p)$ cannot be equal to 0 for any utility function which implies that $\max_{p\in S}f(p)$ must be non-zero for any utility function because $\max_{p\in D}f(p) = \max_{p\in S}f(p)$ for the solution set $S$.

This implies that for any utility function $f_i$ in any $F_i$ there exists a point, $p_j$, such that $f_i(p_j)$ is non-zero. This implies that $p_j$ corresponds to a set $t_j$ that includes the $i$-th element of $U$. Since existence of a point, $p_j$, in $S$ such that $f_i(p_j)$ holds for all $i$, $1 \leq i\leq \left\vert U\right\vert$, then there exist corresponding sets, $t_j$, in $T$ such that $t_j$ contains the $i$-th element of $U$ for all values of $i$. So, as $T^*$ is the set containing the sets $t_j$ for all values of $j$, then, $T^*$ is a set cover of $U$.

\hfill\ensuremath{\square}

With this, we proceed to prove the correctness of the reduction, $\mathcal{R}$.

\begin{lemma} \label{reductionCorrectness}
	Correctness of $\mathcal{R}$. An instance of the SC problem, ${I_{SC}}$, has a set cover of size at most $k$ if an only if the corresponding instance of the FAM, ${I_{FAM}}$, has a solution with average regret ratio of 0.
\end{lemma}

\textit{Proof.} Based on the reduction, the instance ${I_{FAM}}$ finds a set $S$ of size $k_{FAM}$ so that the average regret ratio of the set is the lowest possible. Based on Lemma \ref{redCorr2}, the set $S$ corresponds to a set cover of size $k$ of $U$ in the ${I_{SC}}$ instance if and only if its average regret ratio is 0. Because the problem is solved optimally, if the average regret ratio is not zero, there is no set cover of size $k$ or less. If the average regret ratio is zero, $S$ itself corresponds to a set cover of of size $k$.

\hfill\ensuremath{\square}

As a result of this lemma, we can claim that FAM is NP-hard as follows. As shown in Lemma \ref{reductionCorrectness}, An instance of the SC problem, ${I_{SC}}$, has a set cover of size at most $k$ if an only if the corresponding instance of FAM created by the reduction $\mathcal{R}$, ${I_{FAM}}$, has a solution with average regret ratio 0. Furthermore, based on Lemma \ref{redPol}, $\mathcal{R}$ runs in polynomial-time (Note that the average regret ratio of the solution set can be calculated in polynomial time as well). This implies that there exists a polynomial-time reduction, namely $\mathcal{R}$, from SC to FAM such that if we can solve FAM optimally, we can answer the problem SC. Since SC is NP-complete, FAM is NP-hard.

\hfill\ensuremath{\square}

\smallskip
\noindent
\textit {Proof of Theorem~\ref{thm:supermodularity}}. Note that from the definition of supermodularity, we have to show that for all $S \subseteq T \subseteq D$ and for any element $p\in D - T$, $arr(S \cup \{p\}) - arr(S) \leq arr(T \cup \{p\}) - arr(T)$. To do so, consider an element $p \in D - T$. There are two possibilities depending on whether $p$ is the best point in $S$ for any user or not.

\textit {Case 1:} There does not exist any user for whom $p$ is the best point in $S$. That is, $\forall f \in F$, $\exists q \in S$, $f(q) \geq f(p)$. Therefore, for any utility function $f, sat(S \cup \{p\}, f) = sat(S, f)$ and as a result, the regret ratio of none of the users will change, which implies that the average regret ratio will remain unaffected. As such, $ arr(S \cup \{p\}) - arr(S) = 0$. Note that $S \subseteq T$. So, all the points that are the best point of the users in $S$ will exist in $T$ and $p$ will not be the best point for any user in $T$ either. Therefore, $arr(T \cup \{p\}) - arr(T) = 0$. Hence, in this case $arr(S \cup \{p\}) - arr(S) = arr(T \cup \{p\}) - arr(T)$, which completes the proof.

\textit {Case 2:} Assume that there exists at least a user whose satisfaction will increase if we add the point $p$ to $S$. Let $U$ be the set containing such users. Note that for any such user $u \in U$, we have $sat(S \cup \{p\}) = u(p)$. Then, by the definition of the average regret ratio, we have
\begin{multline*} 
arr(S \cup \{p\}) = \int_{f\in F, f \notin U} rr(S \cup \{p\}, f)\eta (f)df +\\
 \int_{f \in U} rr(S \cup \{p\}, f)\eta (f)df
\end{multline*}

In first line of the above equation, we can replace $rr(S \cup \{p\}, f)$ by $rr(S, f)$, since for all $f\in F -  U$ adding $p$ to $S$ does not increase the satisfaction of any user. Note that $rr(S \cup \{p\}, f) = \frac{sat(D, f) - sat(S, f)}{sat(D, f)} - \frac{sat(S \cup \{p\}, f) - sat(S, f)}{sat(D, f)}$. Hence, $rr(S \cup \{p\}, f) = rr(S, f) - \frac{sat(S \cup \{p\}, f) - sat(S, f)}{sat(D, f)}$. Then, we can rewrite the equation as
\begin{multline}\label{IntegralUnion2}
arr(S \cup \{p\}) = \\
\int_{f\in F, f\notin U} rr(S, f)\eta (f)df + \int_{f \in U} rr(S, f)\eta (f)df\\
 - \int_{f \in U}\frac{sat(S \cup \{p\}, f) - sat(S, f)}{sat(D, f)}\eta (f)df
\end{multline}

Let $x$ be a positive constant equal to $\int_{f \in U}\frac{sat(S \cup \{p\}, f) - sat(S, f)}{sat(D, f)}\eta (f)df$. $x$ is positive because for any utility function $f$, $sat(S \cup \{p\}, f) \geq sat(S, f)$. Then the equation can be simplified as $arr(S \cup \{p\}) - arr(S) = -x$.

Now, we add the point $p$ to the set $T$. If the best point of none of the users in $T \cup \{p\}$ is equal to $p$, then we are done. Because $arr(T \cup \{p\}) - arr(T) = 0$, but $arr(S \cup \{p\}) - arr(S) \leq 0$.

Otherwise, the best point of a user $u \in U$ in $T \cup \{p\}$ is different from his or her best point in $T$, i.e. the best point becomes the point $p$. Note that if the best point of a user changes to $p$ when $p$ is added to $T$, then that user must be in $U$. This is because for a user $u \in F$, the change of his or her best point means $u(p) > \max_{q \in T}u(q)$. Since $S \subseteq T$, $u(p) > \max_{q \in S}u(q)$. This means that $u(p) > sat(S, u)$. Thus, addition of $p$ to $S$ changes the best point of the user $u$ in $S$.

Moreover, $sat(T, u)$ can never be less than $sat(S, u)$, because $S$ is a subset of $T$. Thus, $sat(T, u) \geq sat(S, u)$.

Following the same procedure as $(\ref{IntegralUnion2})$ but replacing $S$ and $x$ with $T$ and $y$ respectively, we get $arr(T \cup \{p\}) - arr(T) = - y$, where $y = \int_{f \in U}\frac{sat(T \cup \{p\}, f) - sat(T, f)}{sat(D, f)}\eta (f)df$. Hence, we have

\begin{equation}\label{xMy}
\begin{split}
x - y = \int_{f \in U}\frac{u(p) - sat(S, u) - u(p) + sat(T, u)}{sat(D, f)}\eta (f)df\\
 = \int_{f \in U}\frac{sat(T, u) - sat(S, u)}{sat(D, f)}\eta (f)df
\end{split}
\end{equation}

Since $sat(T, u) \geq sat(S, u)$, then $x \geq y$, which implies that $arr(S \cup \{p\}) - arr(S) \leq  arr(T \cup \{p\}) - arr(T)$. This completes the proof.

\hfill\ensuremath{\square}

\smallskip\noindent
\textit {Proof of Lemma~\ref{thm:monotonicallyNonIncreasingFunction}}. To show that $arr(S)$ is monotonically decreasing, we need to show that for all $S \subseteq U$ and $p \in U$, $arr(S \cup \{p\}) \leq arr(S)$, or equivalently, $arr(S \cup \{p\}) - arr(S) \leq 0$. The proof of Lemma~\ref{thm:monotonicallyNonIncreasingFunction} follows a structure similar to the proof of Theorem~\ref{thm:supermodularity}. Two cases are possible depending on whether $p$ is the best point in $S$ for any user or not.

\textit {Case 1:} There does not exist any user for whom $p$ is the best point in $S$ and satisfaction of no user will increase if we add the point $p$ to $S$. Therefore, for any utility function $f, sat(S \cup \{p\}, f) = sat(S, f)$ and the regret ratio of all of the users will not change with the addition of the point, which means that the average regret ratio will remain unaffected. As such, $arr(S \cup \{p\}) - arr(S) = 0$ which completes the proof in this case.

\textit {Case 2:} Assume that there exists at least a user whose satisfaction will increase if we add the point $p$ to $S$. Let $U$ be the set containing such users. Then, based on the equation $(\ref{IntegralUnion2})$ proven in Theorem~\ref{thm:supermodularity}, $arr(S \cup \{p\}) - arr(S) = -x$ where $x = \int_{f \in U}\frac{sat(S \cup \{p\}, f) - sat(S, f)}{sat(D, f)}\eta (f)df$. Users in $U$ are selected such that $sat(S \cup \{p\}, u) - sat(S, u) \geq 0$, which implies that $x \geq 0$. Thus, $arr(S \cup \{p\}) - arr(S) \leq 0$ which proves the theorem.

\hfill\ensuremath{\square}

\textit{Proof of Theorem~\ref{thm:steepness}}. Based on \cite{Vitorpi:supermodular}, minimizing a monotonically decreasing supermodular function with the steepness $s$, using an algorithm that in each iteration removes a point whose removal increases the value of the function the least, will result in a solution which is within $\frac{e^t - 1}{t}$ factor of the optimal solution, where $t = \frac{s}{1-s}$. Based on Theorem ~\ref{thm:supermodularity} and Lemma ~\ref{thm:monotonicallyNonIncreasingFunction}, $arr(S)$ is a monotonically decreasing supermodular function. Since Algorithm $\ref{algo}$ follows the same procedure as \cite{Vitorpi:supermodular}, Algorithm $\ref{algo}$ will return a solution with approximation ratio $\frac{e^t - 1}{t}$ of the optimal solution.

\hfill\ensuremath{\square}

\textit{Proof of Lemma~\ref{lemma:lowerBound}:} 
Algorithm \ref{algo:greedy} removes one point from the solution set at the end of each iteration. Hence, $S_{curr} \subseteq S_{prev}$ and for each point $p \in S_{prev}$, $S_{curr} - p$ is a subset of $S_{prev} - p$ while they differ in at most one point. Since the average regret ratio is a monotonically decreasing set function, $arr(S_{prev} - p) \leq arr(S_{curr} - p)$, or, $v_{p, S_{curr}}\geq v_{p, S_{prev}}$.

\hfill\ensuremath{\square}

\textit{Proof of Lemma~\ref{lemma:lowerBoundUsage}:} 
We aim at finding a point $p_o = \arg\min_{p\in S_{curr}}v_{p, S_{curr}}$. Consider a point $p$ in $S_{curr}$ and let $v$ be the evaluation value of $p$ based on $S_{curr}$. If a point $p'$ has an evaluation value based on $S_{prev}$ larger than $v$, i.e., $v_{p', S_{prev}} > v$, then by Lemma~\ref{lemma:lowerBound} we have $v_{p', S_{curr}} \geq v_{p', S_{prev}} > v$. Thus, $p'$ has an evaluation value larger than that of $p$ and thus does not have the minimum evaluation value, i.e., $p' \neq p_o$.
	
If there is no point in $S_{curr}$ whose evaluation value based on $S_{prev}$ is smaller than $v$, we have $\forall p' \in S_{prev}, v_{p', S_{prev}} \geq v$. By Lemma~\ref{lemma:lowerBound} we have $\forall p' \in S_{prev}, v_{p', S_{curr}} \geq v_{p', S_{prev}} \geq v$. Hence $p$ has the lowest evaluation value in $S_{curr}$, i.e. $p = p_o$.

\hfill\ensuremath{\square}

\smallskip\noindent
\textit{Proof of Theorem~\ref{thm:samplingTheory}}. The proof uses Chernoff bounds \cite{wong:kHit} which are inequalities stated as follows.

\indent \textit{Chernoff bound.} Assume $X_1, ..., X_N$ are N independent random variables, where $0 \leq X_i \leq 1$ for all $i$. Let $X = X_1 + X_2 + ... + X_N$ and $\mu = E[\,X]\,$. Then, for any $\epsilon\geq 0$, $ Pr[\,X \geq (\,1 + \epsilon)\,\mu]\, \leq e^{\frac{-\epsilon^2}{2 + \epsilon}\mu}$ and 
$Pr[\,X \leq (\,1 - \epsilon)\,\mu]\, \leq e^{\frac{-\epsilon^2}{2}\mu}$.

Let $X_i$  be a random variable denoting the regret ratio of the $i_{th}$ randomly selected utility function and $F_N$ be the set of all the randomly selected utility functions. $X_i$ maps $F$, to $R_{\geq 0}$ and takes the value $rr(S, f)$ where $f \in F$ with the probability $\eta(f)$. So, for a randomly selected utility function $f$, $X_i$ denotes $\frac{\max_{p\in D}f(p) - \max_{p\in S}f(p)}{\max_{p\in D}f(p)}$, where $S$ is the set of $k$ selected points shown to the user. Then, $E[\,X_i]\,$ for all $i$, $1 \leq i \leq N$ is equal to $arr^*$. By definition, we also have that the average regret ratio of the $N$ randomly selected users can be written as $arr = \frac{\sum_{i = 1}^{N} X_i}{N}$.

X and $\mu$ are defined in the same way as in the Chernoff bound. Then $\mu = E[\,X]\, = \sum_{i = 1}^{N} E[\,X_i]\,$ by linearity of expectation. Since $X = \sum_{i = 1}^{N} X_i$, we have $X = N\times arr$. Since $E[\,X_i]\,$, is $arr^*$, then $\mu = \sum_{i = 1}^{N}arr^* = N\times arr^*$. By applying the first equation of the Chernoff bounds, we get $Pr[\,X - \mu \geq \epsilon'\mu]\, \leq e^{\frac{-\epsilon'^2}{2 + \epsilon'}\mu}
 \leq e^{\frac{-\epsilon'^2}{3}\mu}$.

Now, let $\epsilon'' = \epsilon'\mu$. We obtain $Pr[\,X - \mu \geq \epsilon'']\,  \leq e^{\frac{-\epsilon''^2}{3\mu}}$. Also, let $\sigma = e^{\frac{-\epsilon''^2}{3\mu}}$. Then we have $\epsilon'' = \sqrt{3\mu ln \frac{1}{\sigma}}$ By substituting the value of $\epsilon''$ into the previous inequality, we obtain

\begin{equation} \label{main1}
Pr[\,X - \mu \geq \sqrt{3\mu ln \frac{1}{\sigma}}]\,  \leq \sigma
\end{equation}

As mentioned before, $\mu = N\times arr^*$ and $X = N \times arr$. Substituting these into $(\ref{main1})$ we get $Pr[\, arr - arr^* \geq \sqrt{\frac{3(\,arr^*)\, ln \frac{1}{\sigma}}{N}}]\,  \leq \sigma$.

But since $arr^* \leq 1$, we have that $Pr[\, arr - arr^* \geq \sqrt{\frac{3ln \frac{1}{\sigma}}{N}}]\,  \leq \sigma$.

If we let $\epsilon$ be $\sqrt{\frac{3ln \frac{1}{\sigma}}{N}}$ then we have $Pr[\, arr - arr^* \geq \epsilon]\,  \leq \sigma$ which implies $Pr[\, arr - arr^* < \epsilon]\,  \geq 1 - \sigma$

An analogous procedure using the second Chernoff inequality will result in $Pr[\, arr^* - arr < \epsilon]\,  \geq 1 - \sigma$ which together with the last inequality above implies that $\left\vert arr - arr^* \right\vert < \epsilon$ with the confidence of at least $1 - \sigma$ when the sample size is $N = \frac{3ln \frac{1}{\sigma}}{\epsilon^2}$

$\hfill\ensuremath{\square}$

\textit {Proof of Theorem~\ref{thm:discreteDistribution}}. It follows from Theorems~\ref{thm:steepness} and ~\ref{thm:samplingTheory}.

\hfill\ensuremath{\square}

\fi

\end{sloppy}
\end{document}